# Scan-Coil Delay Causes Anisotropic Signal Loss in Fast 4D-STEM

Vishal Kumar[1,2,#], Andreas Jehle,[3] Tizian Lorenzen,[3] Julika Radecke[1,2], Knut Müller-Caspary,[3] Massimo Kube[1,2], and Henning Stahlberg[1,2,#]

1. Laboratory of Biological Electron Microscopy, Institute of Physics, School of Basic Sciences, EPFL, Rt. de la Sorge, 1015 Lausanne, Switzerland
2. Department of Fundamental Microbiology, Faculty of Biology and Medicine, University of Lausanne, Rt. de la Sorge, 1015 Lausanne, Switzerland.
3. Department of Chemistry and Center for NanoScience, Ludwig-Maximilians-Universität München, Butenandtstr. 11, 81377 München, Germany

\# Corresponding authors: Vishal.Kumar@epfl.ch, Henning.Stahlberg@epfl.ch

## Abstract

Fast pixelated detectors are driving 4D-STEM toward microsecond dwell times, a regime in which the finite response of the scan deflection coils becomes comparable to the dwell time itself. Using direct probe imaging and sub-frame diffraction analysis, we document a significant intra-dwell scan-coil delay that systematically smears the recorded signal anisotropically along the fast scan direction, with a settling timescale of several tens of microseconds. We present a phase-correlation-based sub-frame alignment procedure that measures and corrects this smearing, and we assess its impact on focused and defocused 4D-STEM reconstructions over a range of scan step sizes. The correction restores signal across a broad range of spatial frequencies, with the largest gains at the large step sizes required for low-dose biological imaging. Because it operates on existing data with no modification to the microscope, the method offers a practical route to recovering signal that would otherwise be lost to scan-coil delay.

## Highlights

1. High-speed 4D-STEM is limited by finite scan-coil settling delay.
2. Scan-coil settling smears the recorded signal leading to anisotropic information loss
3. This beam smearing can be computationally corrected, recovering much of the lost signal.
4. Recovery spans a broad frequency band, with the largest gains at larger step sizes.

# Graphical abstract

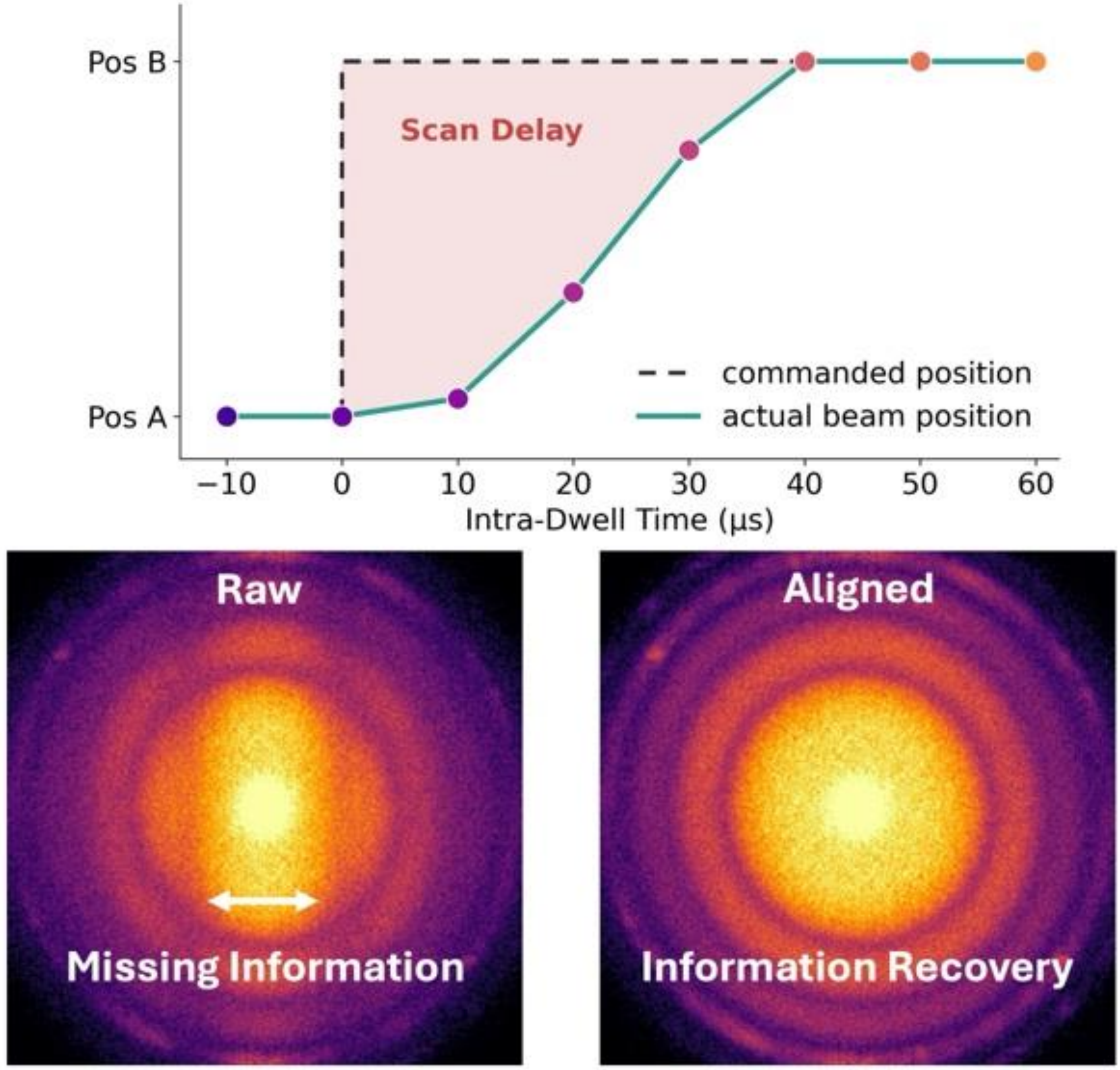

# 1 Introduction

Scanning transmission electron microscopy (STEM) is a widely employed tool for studying the structural and chemical properties of materials at atomic resolution [1–4]. STEM can leverage various imaging modes, such as bright-field (BF), high-angle annular dark field (HAADF) [5,6], and differential phase contrast (DPC) [7,8], by changing detector geometries. Four-dimensional STEM (4D-STEM) extends conventional STEM by recording a full electron scattering pattern (2 diffraction coordinates) at each beam position (2 real-space coordinates), enabling a wider range of image analysis approaches that are inaccessible to conventional single-detector techniques [9–11]. Recording the full forward-scattered momentum distribution allows 4D-STEM to support both conventional imaging via virtual detectors and computational methods such as ptychography using the same dataset.

STEM has long been established as a standard technique for nanoscale characterization of radiation-hardened materials, but its application to beam-sensitive samples has remained limited. Recent advances in high-speed pixelated detectors, combined with computational 4D-STEM approaches, are now closing this gap owing to their low dose requirements and high sensitivity to weakly scattering elements [12,13]. Building on these advances, recent studies have demonstrated atomic-resolution imaging of metal–organic frameworks (MOFs) at 2 Å resolution with electron doses below 100 $e^-/Å^2$ [14], using electron ptychography. Similarly, computational techniques have likewise been applied to biological samples, achieving sub-nanometer resolution at fluences below 40 $e^-/Å^2$ [15,16].

Further progress toward imaging beam-sensitive materials, particularly biological specimens, will require the implementation of dose-fractionation and motion-correction strategies [17–19]. These approaches push imaging into an even lower dose regime, demanding faster detectors and shorter dwell times than currently standard. For instance, imaging a biological specimen at a total fluence of 10 $e/Å^2$ with a probe current of 10 pA and a scan step of 2 Å requires a dwell time of ~0.64 µs. To meet these requirements, next-generation detectors such as the Dectris ARINA [20] and event-driven cameras such as TimePix3/4 [21,22] have already been developed and are now being evaluated for low-dose 4D-STEM. However, operating at such low electron doses in 4D-STEM pushes data collection strategies to employing very short dwell-times, which exposes limitations of the scan hardware that were negligible at slower acquisition speeds, and that remain poorly characterized.

A particularly important effect in this faster scanning regime (a few microseconds dwell time) is the scan coil delay itself. The deflection coils are electromagnets of finite inductance, producing a non-zero delay between the application of the driving voltage and the electron beam (a.k.a., the probe) reaching its actual commanded position. The positional consequences of this lag have been studied extensively. The finite scan-coil response is well known to deform the first few pixels of each scan line [23,24], an effect typically mitigated by inserting a flyback delay of 100-1000 µs at the start of every line [25], while post-hoc strategies such as flyback hysteresis correction [26,27], non-raster (*e.g.*, spiral) scan paths [28,29], and ptychography-specific scan-position refinement methods [30,31] can recover accurate probe coordinates after acquisition. These approaches all share the same underlying assumption that each recorded diffraction pattern corresponds to a single, potentially mislocated probe position.

This assumption breaks down once dwell times approach the coil response time. In that regime, the probe is not merely displaced during the dwell window. Rather, the probe remains in motion, traversing a finite trajectory across the sample while the detector is integrating. Each diffraction pattern therefore captures a continuous superposition of patterns from neighboring positions rather than a single coordinate, smearing both the real-space registration and the diffraction signal itself. This intra-dwell smearing has received very limited attention in the literature, simply because it was mostly inaccessible due to limited cameras frame rates (~1000 fps). This effect has only become directly measurable with the current generation of ultrafast cameras [32], such as the Dectris ARINA and TimePix3/4. In principle the smearing can be avoided in hardware by the use electrostatic or radio-frequency beam-blanking [33–35], which can deflect the beam on nanosecond timescales, so that the specimen is not illuminated while the probe is still settling. However, both of these solutions require intensive modification of the instrument.

The effect of probe smearing is especially pronounced for biological specimens, where large step sizes (~10 Å) are typically used to cover wide fields of view and to minimize cumulative electron dose [15,16]. Under the required conditions of larger step sizes and fast data collection, the probe drags across a substantial portion of each step before settling, smearing the recorded signal over a much larger area than intended, which ultimately limits the quality of reconstructed images.

In this study, we characterize the influence of intra-dwell scan-coil smearing in fast 4D-STEM. Using direct probe imaging and sub-frame diffraction analysis, in which each scan step is recorded as a series of shorter exposures (sub-frames) rather than a single exposure, we show that the scan coils continue to settle for tens of microseconds after each commanded position jump, smearing the recorded signal along the fast-scanning direction. We then introduce a phase-correlation based sub-frame alignment that recovers the signal that was lost to this smearing across a range of reconstruction modes and scan step sizes. These results open a pathway for fast-frame 4D-STEM to reach its full potential, as detector technologies continue to push toward shorter dwell times, particularly for low-dose imaging of beam-sensitive and biological specimens.

# 2 Methods

**2.1 Microscope and detector:** 4D-STEM data were acquired on a Thermo Fisher Scientific Titan Krios G4 (at EPFL) transmission electron microscope equipped with a cold field-emission gun (cold-FEG), operated at 300 kV with a probe current of 95 pA and a semi-convergence angle of 24 mrad. Probe-corrector alignment was performed before each acquisition using the S-CORR software (Thermo Fisher Scientific). Scan control was provided by a TVIPS Universal Scan Generator (USG), which allowed the deflection coils to be driven with arbitrary scan trajectories. Diffraction patterns were recorded with a Dectris ARINA pixelated detector operated at a 100 kHz frame rate in 96 × 96 two-fold binned mode, with a 10 µs sub-frame integration window. A subset of measurements was performed on a second instrument, an aberration-corrected FEI Titan Themis (S)TEM (60-300 kV) at Ludwig-Maximilians-Universität München (LMU) using a DigiScan (Gatan) scan generator and the data was recorded using event-driven pixelated detector, TimePix4. The microscope and scan-generator names are listed in **Supplementary Table 1**.

**2.2 Data acquisition:** A standard gold cross-grating reference sample was used for all measurements. For the single-position measurements in **Figure 1**, the measurements were done in real-space rather than diffraction space. The probe semi-convergence angle was 24 mrad, operated at 300 kV at cryo conditions (~ 77 K). The probe was held at a fixed commanded position while the detector streamed consecutive 10 μs sub-frames. For the full 4D-STEM acquisitions, the scan coils were operated at 10 kHz while the detector ran at 100 kHz, yielding ten consecutive 10 μs sub-frames per commanded probe position. Linear, serpentine, and spiral scan patterns were generated within the USG. For all subsequent reconstruction analyses (Figures 4–6), a linear raster scan was used using a TVIPS universal scan-generator, as it is the most commonly employed scan pattern in 4D-STEM. Focused-probe acquisitions (**Figures 4 and 5**) were performed on a 512 × 512 scan grid, and defocused parallax acquisitions (**Figure 6**) on a 256 × 256 scan grid. Scan step sizes are reported in the corresponding figure captions; data are presented for step sizes ranging from 0.32 Å to 5.025 Å. Defocused parallax acquisitions used a probe defocus of approximately 90 nm. For measurements at LMU, microscope was operated at 200 kV with a semi-convergence angle of 21.1 mrad. Diffraction patterns were recorded with a TimePix4 event-driven pixelated detector. Defocused parallax data were acquired on the same gold cross-grating reference sample at a scan step of 2.16 Å and a probe defocus of approximately 60 nm, on a 256 × 256 scan grid with a total dwell time of 25 μs per probe position. The event-driven TimePix4 data were subsequently time-binned into five consecutive sub-frames of 5 μs each. These acquisitions were used to confirm that the intra-dwell scan-coil settling, and its correction are reproduced on an independent microscope and detector **(Supplementary Figures 1 and 12)**.

**2.3 Reconstruction:** Virtual dark-field (vDF) images were computed using a custom Python notebook, with an annular integration window from 40 to 50 mrad, positioned outside the central bright-field disc. Integrated centre-of-mass (iCoM) images and defocused parallax reconstructions were computed with the open-source py4DSTEM package [36]. For all three reconstruction modes, the sub-frame structure of the data was preserved, so that each reconstruction was computed separately for each sub-frame index before any averaging or summation.

**2.4 Sub-frame alignment:** The intra-dwell drift was corrected by aligning the per-sub-frame reconstructed images in 1D along the scan trajectory. For each pair of sub-frames, the 2D reconstructed image was unrolled into a 1D signal along the actual scan path (linear, serpentine, or spiral). The translation between each early sub-frame and the final (settled) sub-frame was then measured by upsampled phase cross-correlation with an upsampling factor of 20 (1/20-pixel precision), using the *phase_cross_correlation* function from *scikit-image* [37]. The measured shift was applied to the corresponding 1D signal, and the result was re-rasterised onto the original 2D scan grid to get the aligned image. A single scalar shift was applied per sub-frame, uniform across the field of view.

**2.5 Recovery factor analysis:** Power spectra of reconstructed images were computed with a Hann window to suppress edge artefacts. The recovery factor was defined as the ratio of the aligned to the raw power spectrum, $R(\mathbf{k}) = P_{\text{aligned}}(\mathbf{k})/P_{\text{raw}}(\mathbf{k})$. For focused-probe reconstructions (**Figures 4 and 5**), the recovery factor was evaluated on narrow annular windows of width ±0.025 $Å^{-1}$, centered at each accessible Au Bragg reflection and plotted as a function of azimuthal angle. For defocused parallax reconstructions (**Figure 6**), it was evaluated as a function of radial spatial frequency, averaged over ±20° wedges centered on the fast-scanning ($k_x$) and slow-scanning ($k_y$) directions.

# 3 Results

To resolve the intra-dwell behavior of the scan coils directly, we acquire data with the probe translating over the sample at a slower pace, while the detector streams sub-frames at 10 times faster speed. This was achieved by operating the scan coils at 10 kHz while the Dectris ARINA camera ran at 100 kHz, so that ten consecutive 10 µs sub-frames were recorded per commanded probe position. Because the deflection coils have finite inductance, the probe does not arrive at each commanded position instantaneously but instead follows a finite trajectory between commanded positions A and B, while the coils settle (**Figure 1a**). As a result, the initial sub-frames capture the probe at slightly different points along this trajectory. Rather than indexing the ten sub-frames by position, we label them by their intra-dwell time, from 0 to 100 µs in 10 µs steps, giving a time-resolved view of the probe trajectory across the first 100 µs of the dwell with 10 µs temporal resolution. The 0 µs sub-frame is the final frame at the previous commanded position, and the jump to the new position is issued at the end of it; the probe should therefore jump to its new commanded position exactly at the beginning of the first 10 µs sub-frame, or slightly before that, and this sub-frame "10 µs" in theory should be recorded from a beam being already at the new beam position, but in practice, the beam is not yet there. We use this scheme to resolve the coil response in both real and diffraction spaces, recording the beam position through real-space imaging at high magnification.

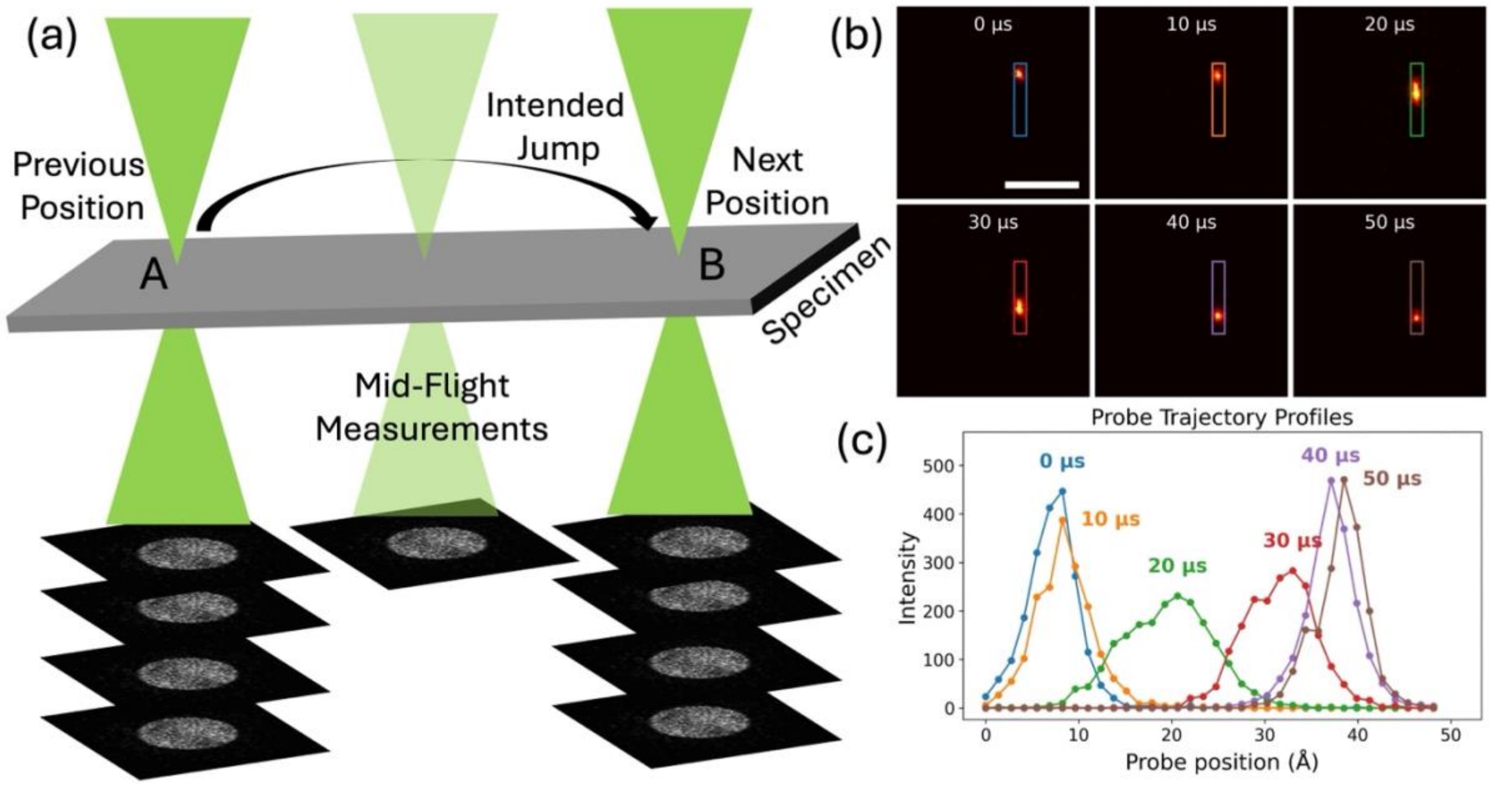

**Figure 1. Intra-dwell smearing in fast 4D-STEM.** (a) At short dwell times, the scan coils cannot move the probe instantaneously between commanded positions A and B (schematic), so mid-flight measurements capture the probe in transit rather than at a single coordinate, smearing the recorded signal. (b) Probe images across a commanded position jump, labelled by intra-dwell time: the 0 µs frame is the final probe image at the previous position (A), and the jump to the new position is commanded at the end of it; the 10–100 µs frames should therefore all show the probe already at the new position (B), but instead the first five (10–50 µs) show the probe still evolving as the coils settle before it stabilizes at B. Scale bar length is 5 nm. (c) Line profiles through each probe in (b), showing the systematic translation and settling dynamics of the probe. Profiles for positions 60 µs to 100 µs are not shown, they are finally indistinguishable from the 50 µs profile.

**Figure 1b** shows ten consecutive sub-frame probe images recorded across one such commanded position jump, using a Titan Krios G4 (at EPFL), equipped with a cold-FEG, Cs probe corrector, and a Dectris ARINA detector. The 0 µs sub-frame is the final frame at the previous position A, and the 10–100 µs sub-frames are those recorded at (or settling toward) the new position B. The new-position deflection command was sent at the end of the 0 µs sub-frame and was synchronized with the camera, so that the 10 µs sub-frame is the first frame recorded after the commanded jump. Tracking the scan-generator output on a high-speed oscilloscope confirmed that this signal switches to the new position quickly, with the jump completing in under a microsecond, so the settling seen in the following sub-frames comes from the coils rather than the drive signal. Rather than appearing fixed at B from the 10 µs sub-frame onwards, the probe is visibly offset from its final location in the early sub-frames and inches toward its commanded position over the following sub-frames, reaching a stable position after ~4 sub-frames (by ~40 µs). The line profiles (**Figure 1c**) clearly show that in the first sub-frame the probe is closer to the previous position A than to the commanded new position B, and the next two sub-frames are progressively displaced along the scan direction, so that the probe intensity is effectively smeared over the first four sub-frames over a length comparable to the scan step. We note that the underlying probe motion is continuous, and what we record in each sub-frame is its integral over a 10 µs detector window. The same behavior was observed across different scan generators and different microscopes: driving the scan with either the external TVIPS Universal Scan Generator (USG) or the internal TFS scan generator and repeating the measurement on a second microscope, a FEI Titan Themis (S)TEM (at LMU Munich), all yielded the same intra-dwell settling of the probe centroid, which converges to the new commanded position within a few tens of microseconds **(Supplementary Figure 1)**. Notably, the settling on the Titan Themis microscope was faster (~20 µs) than on the Titan Krios microscope (~40 µs), which may reflect differences in the deflection-coil material between the two instruments.

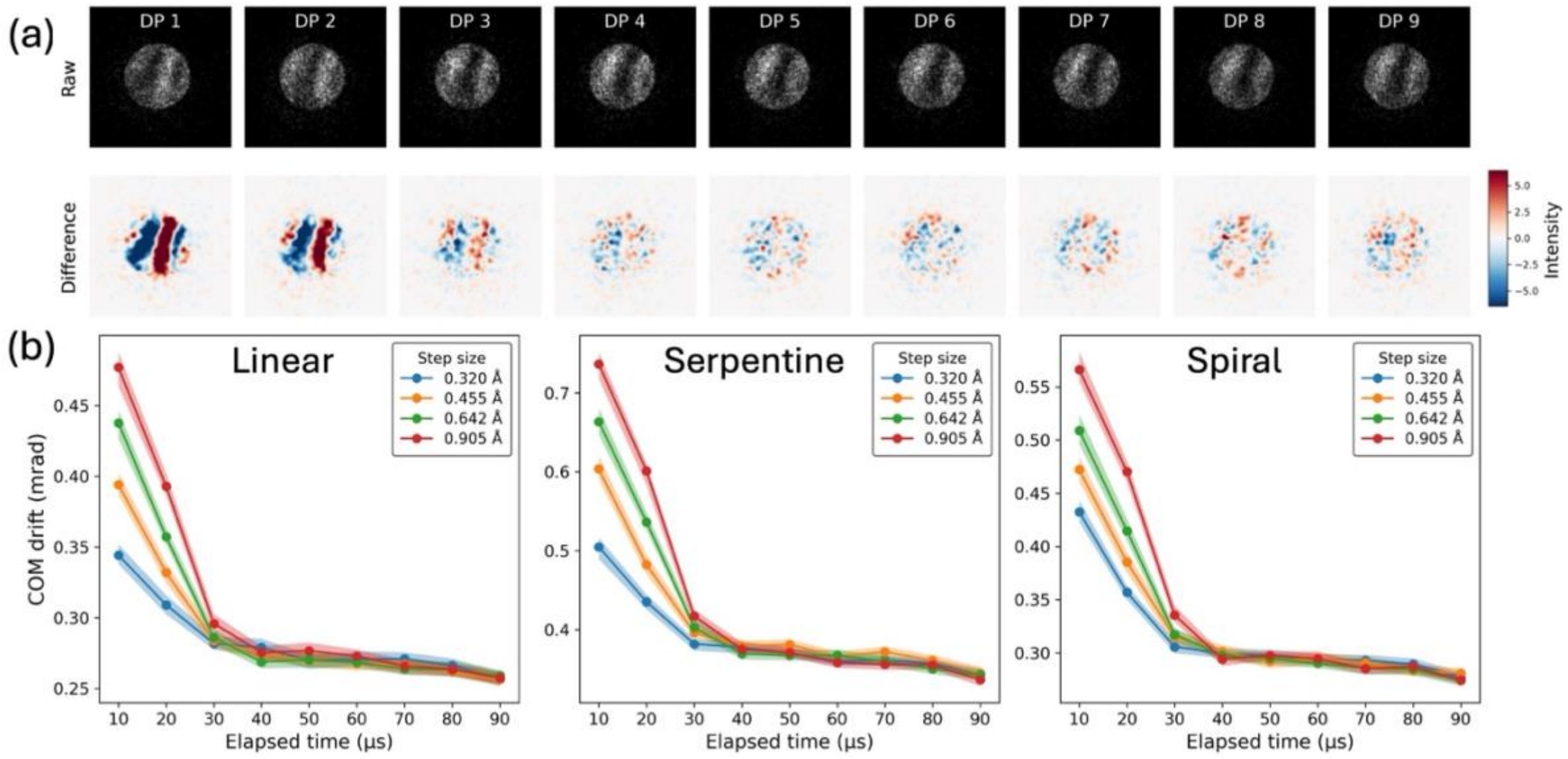

**Figure 2. Direct measurement of intra-dwell CoM drift on a Titan Krios G4.** (a) Nine consecutive 10 μs sub-frame diffraction patterns within a single beam-position (top), and their differences from the final sub-frame (bottom, Gaussian-smoothed with σ = 1.2 pixels for clarity); early sub-frames show systematic intensity redistribution that decays to noise. (b) Median CoM drift versus intra-dwell time for linear, serpentine, and spiral scan strategies at four step sizes (0.32–0.905 Å). Each point is obtained from 6,000 probe positions. The initial drift decays within ~40 μs across all strategies, with larger step sizes producing larger initial drift. Shaded bands represent 95% confidence intervals on the median.

To quantify the effect of this intra-dwell probe motion on the diffraction signal, we acquired a full 4D-STEM scan on a gold cross-grating with 10 sub-frames recorded per probe position, again using the same 10 kHz (scanning speed) and 100 kHz (detector speed) configuration. **Figure 2a** shows nine consecutive 10 μs sub-frame diffraction patterns from a single representative probe position within this dataset, together with their differences from the final (10th) sub-frame. The early sub-frames show a systematic intensity redistribution of the bright-field disc that decays into noise by the fourth sub-frame, mirroring the motion of the real-space probe shown in Figure 1, which also settles by the 4th sub-frame. The same behavior is observed at other positions across the field of view (**Supplementary Figure 2**).

To establish that this is a generic feature of the scan hardware rather than a peculiarity of a single scan pattern or step size, we repeated the acquisition with the TVIPS USG, programming linear, serpentine, and spiral scan strategies at different step sizes (0.32, 0.455, 0.642 and 0.905 Å). **Figure 2b** reports the median Center of Mass (CoM) shift of the bright-field disc as a function of intra-dwell time, taken across ~6000 probe positions in each dataset. All nine combinations show the same qualitative behavior, with an initial drift that decays within roughly 40 μs. The magnitude of the initial drift scales with step size, as expected for a finite-inductance system in which the coils must traverse a larger displacement per commanded jump, but the settling timescale itself is essentially independent of step size and of scan strategy. The independence from scan strategy is particularly informative, as it indicates that the effect is set by the response of the deflection coils and not by the trajectory through which they are driven.

A complete solution to intra-dwell coil drift would ultimately need to be implemented at the hardware level, since the signal recorded in each sub-frame has already been integrated over the probe motion and the lost information cannot be fully recovered after acquisition. Hardware-level mitigation, however, would require either a redesigned low-inductance beam deflection system in the electron microscope or active waveform pre-emphasis on existing instruments [33,35], both of which are expensive and time-consuming to deploy. Here we instead present a software-based correction that operates on data acquired with existing scan hardware and recovers a substantial fraction of the signal that would otherwise be lost to misregistration between sub-frames.

The correction exploits the fact that intra-dwell drift is a temporal offset in the scan sequence rather than a spatial distortion of the image. Because the coil executes the same jump at every scan position, each sub-frame is an image of the same field of view sampled on a rigidly offset scan grid, and the cross-correlation measures that offset directly rather than assuming the probe position. From the 4D dataset, we first form one virtual dark-field (vDF) image per sub-frame index, so that sub-frames 1 through 10 yield ten vDF images of the same field of view. Each vDF image is then unrolled into a 1D signal along the actual scan trajectory (linear, serpentine, or spiral), so that pixels acquired at consecutive moments in time are adjacent in the 1D signal. In this representation, the intra-dwell drift manifests as a single scalar translation of one 1D signal relative to another, and we measure this translation by upsampled phase cross-correlation (1/20-pixel precision) between each early sub-frame and the final, settled sub-frame. The measured shift is applied to the corresponding 1D signal as a sub-pixel translation using Fourier phase shifting, and the result is re-rasterised onto the 2D scan grid (linear, serpentine, or spiral). The procedure yields one scalar shift per sub-frame, applied uniformly across the whole field of view, and is identical for all three scan patterns. The measured shifts follow the scan trajectory rather than a fixed direction, ruling out sample or stage drift, which would produce a translation along a single direction independent of the scan pattern.

The misalignment vectors for the vDF images of sub-frames 1, 2, and 3, relative to the final (10th) sub-frame for the spiral scan-pattern are shown in **Figure 3a**. The systematic drift is directly visible as a translation along the fast-scan direction. **Figure 3b** reports the measured 1D shift as a function of sub-frame index, which decays monotonically toward zero by sub-frame 4, consistent with the real-space and diffraction-space settling timescales in Figures 1 and 2. **Figure 3c** shows the residual shifts after applying the correction and re-running the phase correlation between aligned sub-frames. The residuals are negligible across all sub-frames, confirming that the alignment is exact to within the sub-pixel precision of the registration. The measured shift sequence is also highly reproducible across all three scan strategies, all step sizes, and independent acquisitions. We recover similar per-sub-frame shift values to within the registration precision for raster and serpentine scan-pattern as well (**Supplementary Figure 3 and 4**). This reproducibility further indicates that the intra-dwell drift is a stable property of the scan hardware that can be characterized once, and the same correction can then be applied across different datasets.

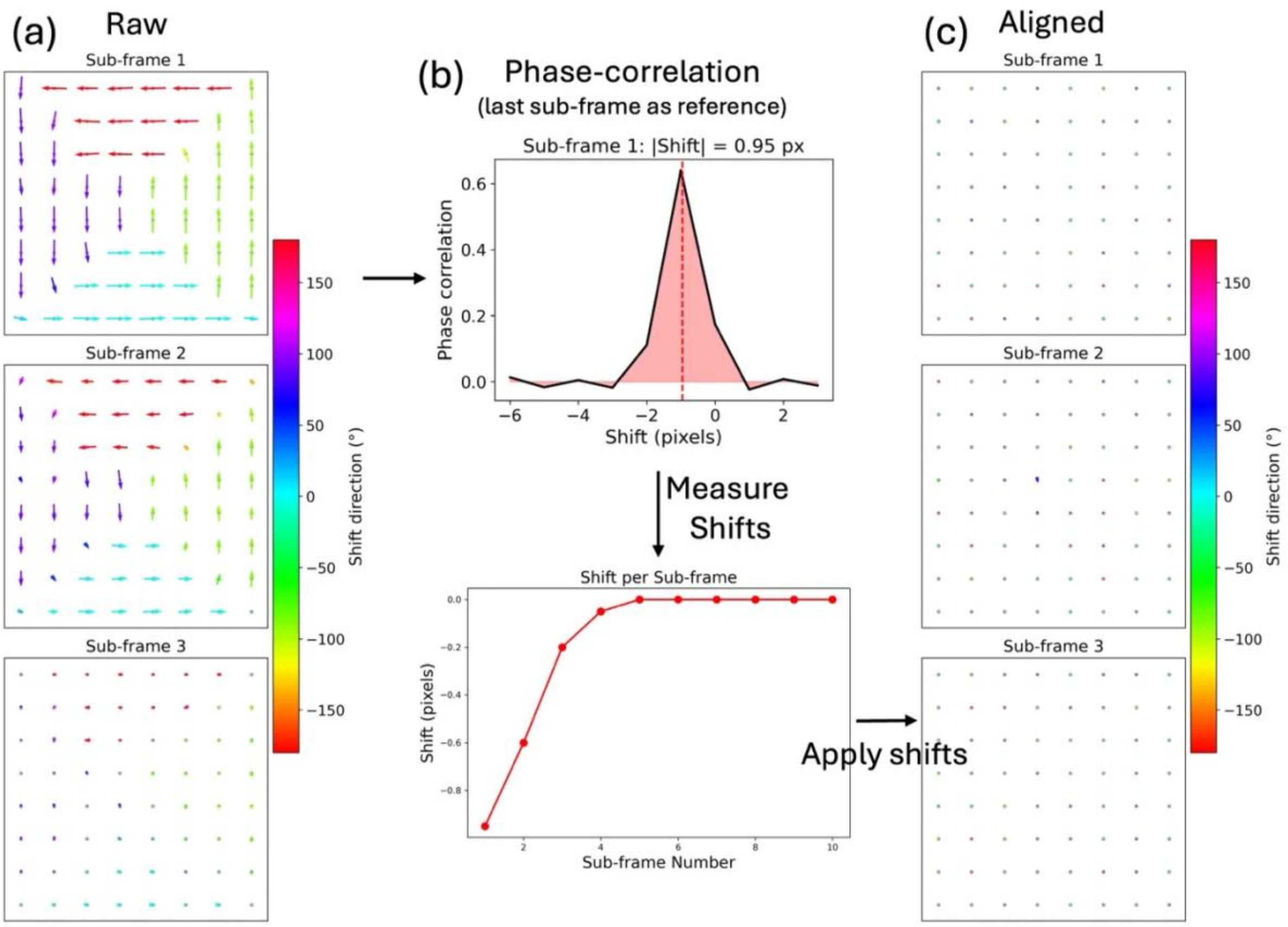

**Figure 3: Sub-frame alignment via phase correlation.** (a) Misalignment vectors for the vDF images of sub-frames 1, 2, and 3, relative to the final (10th) sub-frame for the spiral scan pattern. Each vector shows the measured shift along the fast-scan direction, with arrow length encoding shift magnitude. (b) Phase-correlation peak used to extract the shift (top), and the measured shift as a function of sub-frame index (bottom), which decays as the scan coils settle. (c) Residual shifts after applying the correction and re-running the phase correlation between aligned sub-frames are negligible across all sub-frames.

Having validated the alignment procedure on the raw sub-frames and shown it to be independent of scan strategy (**Figure 3; Supplementary Figures 3 and 4**), we now examine its effect on reconstructed images. From this point on, all reconstructions use a linear raster scan, the most commonly used scan pattern in 4D-STEM. We acquired 4D-STEM datasets with a 50 μs total dwell time, divided into five consecutive 10 μs sub-frames per probe position, on the same gold cross-grating sample. For each dataset we generate two reconstructions, both formed sub-frame by sub-frame and then summed. In the raw image reconstruction, the per-sub-frame images are summed directly, which is equivalent to integrating the full 50 μs dwell. Alternatively, we perform an aligned image reconstruction, in which the measured per-sub-frame shifts are first applied to register the sub-frame images, and the aligned images are then summed. The two reconstructions therefore share the same total integrated signal and differ only in whether the intra-dwell drift has been corrected prior to summation.

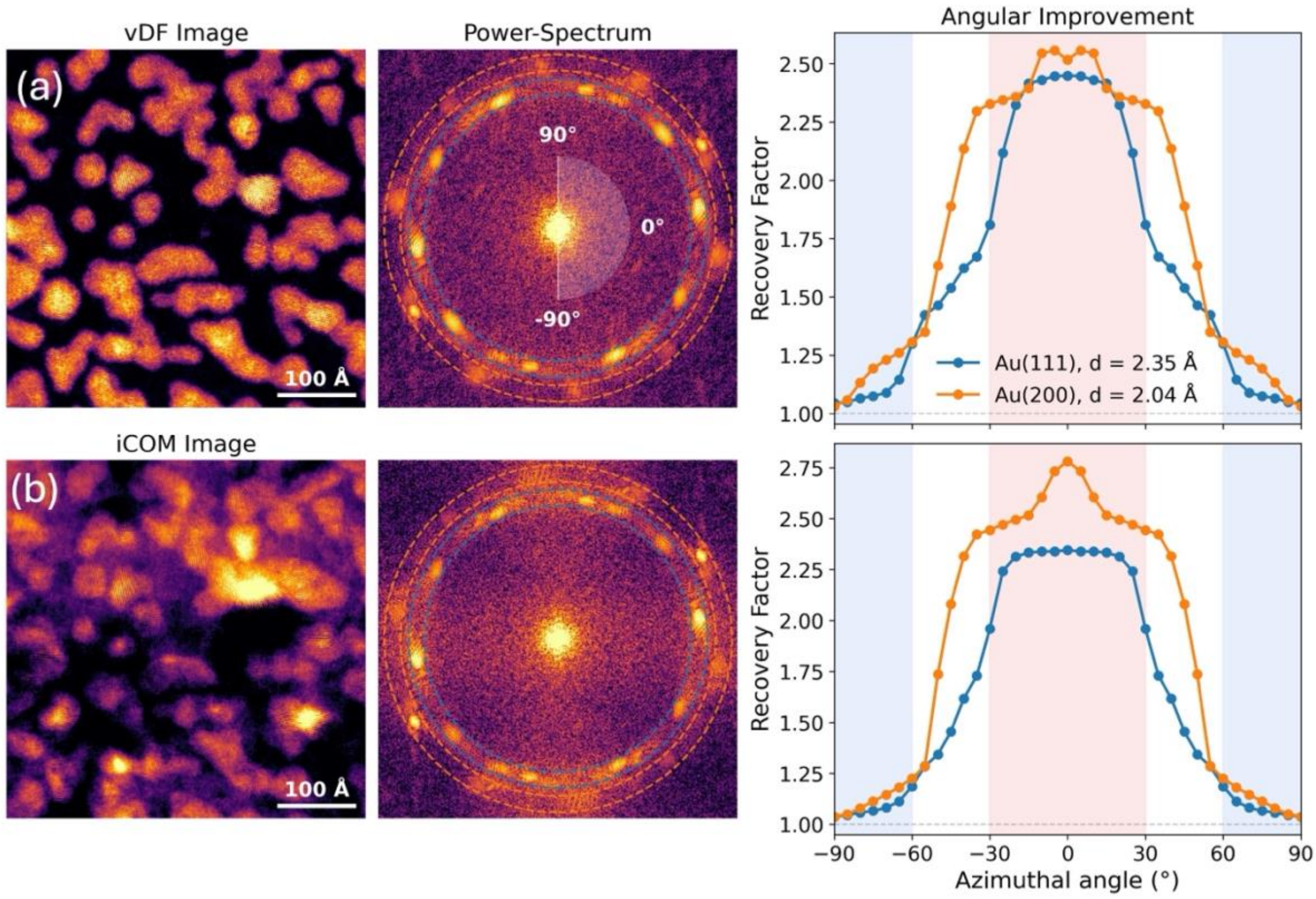

**Figure 4: Comparison of alignment correction across reconstruction modes**. (a) Aligned virtual dark-field (vDF) and (b) aligned integrated centre-of-mass (iCoM) reconstructions of the same 4D-STEM dataset, with corresponding power spectra (middle) and recovery factor versus azimuthal angle (right, the 0° direction corresponds to the fast-scanning direction and 90° is perpendicular to it, i.e. the slow-scanning direction.). Dashed circles on the power spectra mark the integration windows for Au (111) (d = 2.35 Å) and Au (200) (d = 2.04 Å). Pink and blue shaded regions in the azimuthal plots mark the fast-scanning (±30° around 0°) and slow-scanning (±60–90°) directions.

The comparison of the virtual dark-field (vDF) reconstruction (a) and the integrated centre-of-mass (iCoM) reconstruction (b) of the same dataset at a 0.905 Å scan step is shown in **Figure 4**. The two reconstructions are chosen as complementary probes, vDF for amplitude contrast and iCoM for phase contrast, to test whether the intra-dwell smearing acts on amplitude alone or on phase as well. To quantify the effect of the correction on the reconstructed images, we define the recovery factor, ***R(k)*** as the ratio of the aligned to the raw power spectrum,

$$R(k) = \frac{P_{aligned}\ (k)}{P_{raw}\ (k)}$$

integrated over a narrow annular window around each Au reflection (dashed circles in Figure 4, marking Au (111) at d = 2.35 Å and Au (200) at d = 2.04 Å). A recovery factor greater than unity quantifies the fraction of signal that is restored by the alignment at that spatial frequency.

Plotting the recovery factor as a function of azimuthal angle around each annulus reveals a strongly anisotropic recovery profile: in both vDF and iCoM reconstructions, the recovery factor is sharply peaked in the fast-scanning direction (pink shaded region, ±30° around 0°) and remains close to unity in the slow-scanning direction (blue shaded region, ±60–90°). This anisotropy reflects the fact that the smearing acts only along the fast-scanning direction, attenuating the transfer of spatial frequencies along the fast-scan while leaving the perpendicular direction relatively unaffected. Importantly, the vDF and iCoM reconstructions yield indistinguishable angular profiles, indicating that the intra-dwell smearing and its correction act equivalently on amplitude and phase contrast images.

Radial profiles along the fast- and slow-scan directions are shown for both the raw and aligned images in **Supplementary Figures 5** (vDF) and **6** (iCOM). The raw and aligned images appear similar on visual inspection, since the residual displacement is only about one scan pixel and is therefore difficult to discern by eye in a full-field display. The radial profiles resolve it unambiguously, as sub-frame alignment recovers substantial power at the gold lattice spacings along the fast-scan direction, whereas the slow-scan profiles are unchanged. As a further check that the recovered power originates from the unsettled sub-frames, we also formed an image from the settled sub-frames alone (sub-frames 4–5). After normalization the radial profile along fast-scan profile matches that of the aligned image, although it retains only 40% of the total dose (**Supplementary Figure 7**), confirming that the information lost in raw images is confined to the early sub-frames where the beam is still settling down.

We then examine how the recovery factor depends on the scan step size used (0.32 – 0.905 Å), which shows two clear trends (**Figure 5**, an equivalent analysis for the iCoM reconstructions is provided in **Supplementary Figure 8**). Firstly, at any given step size, the recovery factor grows with spatial frequency, with higher-order Au reflections systematically recovering more of the lost signal, than lower-order reflections (**Supplementary Figure 9**). Second, at any given reflection, the recovery factor grows monotonically with step size, reaching nearly 3× at Au (200) for the 0.905 Å step. Both trends are consistent with the intra-dwell drift acting as an effective probe broadening along the fast-scanning direction: larger step sizes correspond to larger absolute smearing lengths, and finer spatial features are more strongly suppressed by a given smearing length. The recovery factor becomes less important when the step size is smaller, but even for our smallest step size of 0.32 Å, the recovery factor stays significant for the highest spatial frequencies even, where it reaches ~1.4× at 1.23 Å for Au (311) spacing. In contrast, in the non-scanning direction, the recovery factor always remains close to unity across all step sizes, as expected, since the smear is not affecting beam positions in the slow-scan direction. This confirms the directional asymmetry established in **Figure 4**. As an independent check, we performed directional Fourier ring correlation (FRC), using an image formed from sub-frames 1–4 as the test and the settled sub-frame 5 as the reference. The resulting resolution improvement along the fast-scan direction follows the same trend with step size, while the slow-scan direction is unaffected (**Supplementary Figure 10**).

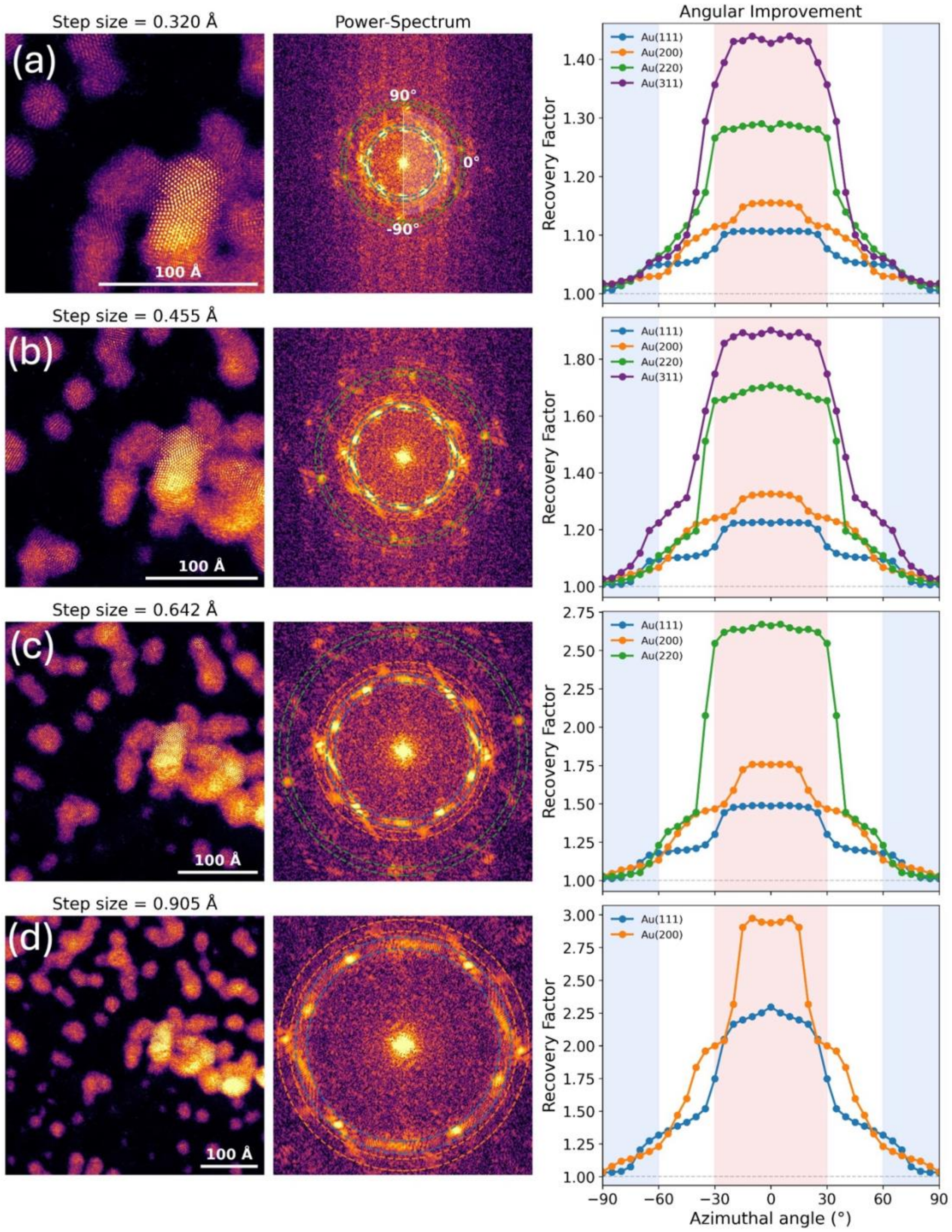

**Figure 5: Step-size and frequency dependence of the alignment correction**. vDF images, power spectra, and recovery factor versus azimuthal angle for four scan step sizes: (a) 0.32 Å, (b) 0.455 Å, (c) 0.642 Å, (d) 0.905 Å. Dashed circles mark integration windows used to compute the recovery factor at each accessible Au reflection (d = 2.35, 2.04, 1.44, 1.23 Å for Au (111), (200), (220), (311)). Pink and blue shaded regions in the azimuthal plots mark the fast-scanning (±30°) and slow-scanning (±60–90°) directions. Higher-order reflections become inaccessible at larger step sizes due to Nyquist sampling.

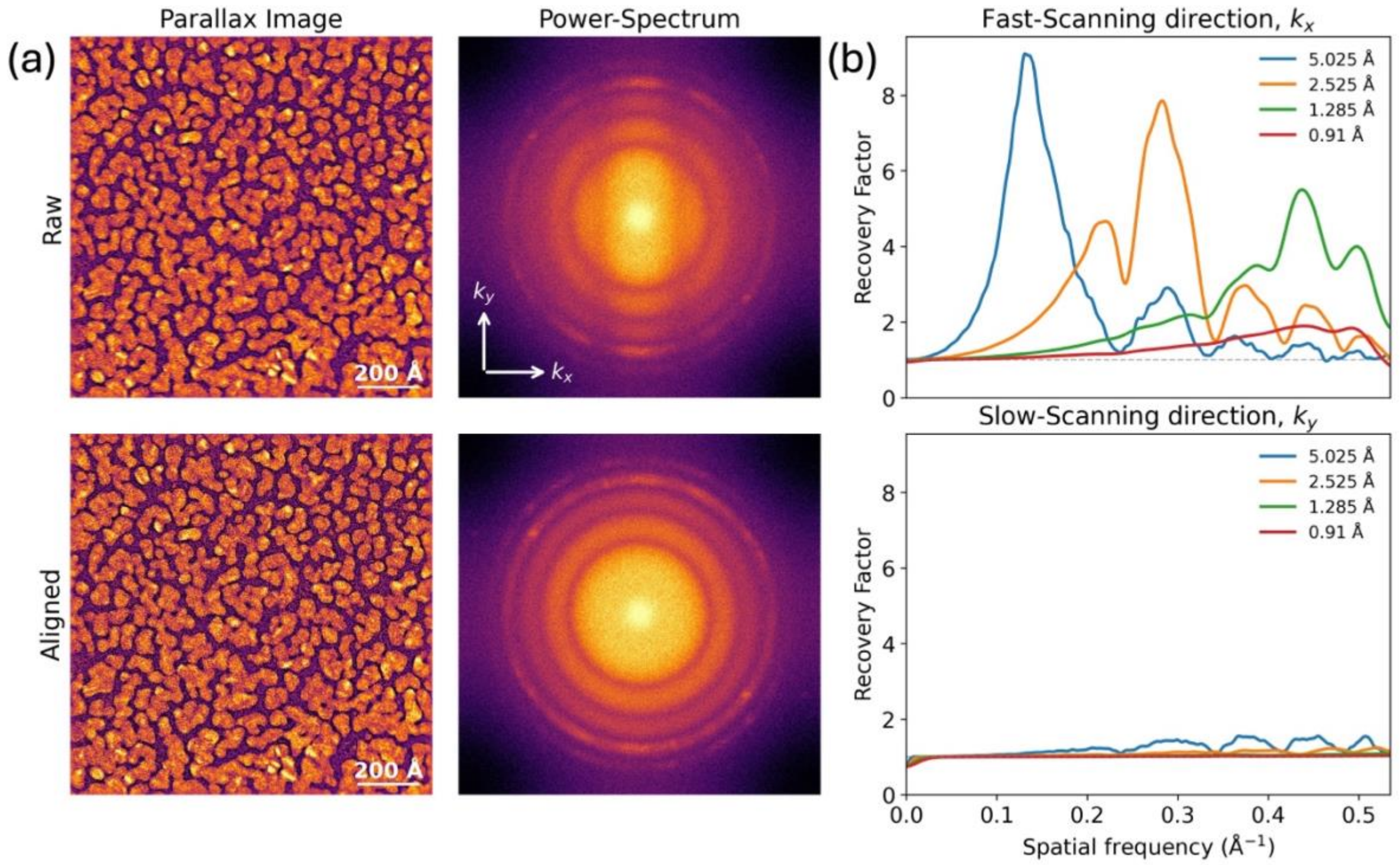

**Figure 6: Broadband recovery in defocused aberration-corrected parallax imaging.** (a) Raw and aligned parallax image for step size of 5.025 Å and its corresponding power spectrum. (b) Recovery Factor versus spatial frequency for four scan step sizes (0.91 – 5.025 Å) in the fast-scanning direction ($k_x$, ±20°) and the slow-scanning direction ($k_y$, ±20°). Curves are Gaussian-smoothed (σ = 3 frequency bins) for clarity.

The focused-probe experiments are limited to relatively small step sizes, since at larger steps the Au Bragg reflections fall outside the Nyquist limit and are no longer accessible in the reconstructed power spectrum. To further evaluate the correction at much larger step sizes, which are useful for low-dose applications such as imaging of biological specimens, we turn to defocused parallax imaging [38] on the same gold cross-grating. In this regime the power-spectrum of the reconstruction is modulated by the oscillating contrast transfer function of parallax imaging,[39] allowing us to evaluate the recovery factor across a continuous range of spatial frequencies. Crucially, although the scan step is larger, the parallax reconstruction is upsampled onto a finer grid than the commanded scan positions, so that spatial frequencies beyond the scan-step Nyquist limit remain accessible and the desired high-frequency information can still be recovered. The upsampling factors used for each step size for our datasets are provided in **Supplementary Table 2**.

**Figure 6a** displays the parallax reconstructions and their power spectra in the raw and aligned images for the step size of 5.025 Å. The raw power spectrum shows vertical bands of attenuated intensity, arising because intra-dwell smearing reduces the transfer of spatial frequencies along the fast-scan direction. The alignment therefore recovers information across a broad and continuous range of spatial frequencies along the fast-scanning direction for parallax reconstructions. This directional recovery is also evident in the radial power-spectrum profiles, where the aligned profile rises well above the raw one along the fast-scanning direction ($k_x$), while the two remain essentially overlapping along the slow-scanning direction ($k_y$) (**Supplementary Figure 11**).

**Figure 6b** extends this analysis to four step sizes (0.91, 1.285, 2.525, and 5.025 Å), plotting the frequency-resolved recovery factor in the fast-scanning ($k_x$) and slow-scanning directions ($k_y$), taken over ±20° wedges. Two trends emerge, first, the magnitude of the recovery factor in the fast-scanning direction grows monotonically with both step size and spatial frequency, mirroring the trend established for the focused-probe case in **Figure 5**, but extending it across a continuous frequency band. Second, the peak of the recovery shifts to lower spatial frequencies as the step size increases, reflecting the fact that the effective smearing length grows with step size and therefore suppresses progressively coarser features. This is especially relevant for imaging of biological specimens, where the large step sizes used in low-dose acquisitions require recovery across both high spatial frequencies (atomic-scale features) and lower frequencies (used for particle picking and alignment) [17,40]. In the non-scanning direction, the recovery factor remains close to unity across all step sizes and frequencies, confirming that the directional asymmetry is preserved throughout the defocused regime as well. Finally, a representative defocused parallax dataset acquired on a second microscope (Titan Themis, 2.16 Å step) shows the same directional recovery **(Supplementary Figure 12)**, with signal restored along the fast-scanning direction and the slow-scanning direction essentially unchanged, confirming that the effect and its correction generalize across instruments.

# 4 Discussion

We here systematically document the influence of intra-dwell scan-coil smearing on fast 4D-STEM data and present a software-based correction that operates on the reconstructed images at the per-sub-frame level. By combining direct probe imaging in real space with sub-frame diffraction analysis across multiple scan strategies, step sizes, and probe-defocus conditions, our results establish three points:

- **The effect is real and generic.** Intra-dwell scan-coil settling is a measurable feature of fast 4D-STEM at the microsecond dwell times, which is now accessible with fast pixelated detectors. Its consistency across linear, serpentine, and spiral scan strategies (**Figure 2**) demonstrates that it is caused by the electromagnetic response of the deflection coils rather than by the chosen scan trajectory.
- **The effect is directional.** Because the coils traverse a finite displacement along the commanded scanning trajectory during each dwell, the smearing acts predominantly along the fast-scanning direction and leaves the perpendicular direction largely unaffected. This signature is observed consistently across all reconstruction modes, step sizes, and probe-defocus conditions examined here.
- **The effect is correctable in software.** The sub-frame alignment procedure introduced in **Figure 3** recovers a substantial fraction of the lost signal in the fast-scanning direction, with the recovery scaling favorably with both step size and spatial frequency.

The implications differ across the application regimes of 4D-STEM. In the materials-science context, where sub-Ångström step sizes are routinely used to achieve atomic-resolution imaging, the recovery factors are modest in absolute magnitude but concentrated at the highest accessible spatial frequencies. Even at our smallest step size (0.32 Å), the recovery reaches ~1.4× at 1.23 Å (for Au (311) spacing), precisely the frequencies of interest in this regime. At the larger scan steps used in defocused 4D-STEM, where the scan step exceeds the reconstructed pixel size of the image, the recovery factors are considerably larger and extend over a much broader range of spatial frequencies, as seen for the 5.025 Å step used for parallax imaging here. This configuration is commonly, though not exclusively, encountered in low-dose imaging of biological and beam-sensitive specimens. This matters because biological 4D-STEM relies on low frequencies for particle picking and alignment, and high frequencies for structural detail. The sub-frame alignment therefore restores signal across both ends of the relevant spatial-frequency range in a single pass, at no cost in dose or acquisition time. Furthermore, the sub-frame structure naturally supports dose-fractionation strategies developed for cryo-EM [17,18], which could provide additional gains for imaging biological specimens with high-resolution.

In conventional STEM imaging with integrating detectors, dwell times of 1–10 µs are shorter than the coil settling time, so the probe never comes to rest, and the scan is effectively continuous. The resulting smear is a stationary blur of one scan step along the fast-scan direction, rather than the partially settled, dwell-dependent displacement characterized here. However, conventional STEM images are highly oversampled (step sizes < 0.2 Å), so the influence of this blurring is much smaller and is often masked by probe shape, scan jitter and specimen drift. The regime of concern is therefore set jointly by the scan step size and by the dwell time relative to the settling time, rather than by either alone.

The software-based correction presented here should not be seen as a replacement for hardware-level solutions to intra-dwell drift, such as low-inductance coil designs or active waveform pre-emphasis, but as a practical complement to them. Other hardware approaches, such as fast electrostatic beam blanking [35,41] or RF-based beam deflection [34] could also suppress the effect at the source by repositioning or gating the beam faster than conventional scan coils allow. The software correction discussed here, applies directly to data acquired on existing instruments and recovers a substantial fraction of the otherwise lost signal, while hardware solutions would reduce the magnitude of the residual the software must handle. The per-sub-frame shifts measured here may also serve as a quantitative diagnostic of coil-settling performance.

Several aspects of the present correction merit further development. The temporal resolution of the alignment is fundamentally limited by the detector frame rate, so that the underlying continuous probe motion is integrated over each 10 µs sub-frame and a residual smearing therefore persists within every sub-frame, particularly the earliest ones where the coils are moving fastest. Event-based pixelated detectors, which timestamp individual electron events at the nanosecond level [21], could in principle eliminate this residual by allowing the data to be re-binned post hoc into much shorter temporal slices, at the cost of more demanding data handling and lower signal-to-noise ratio. Additionally, given the fact that the settling trajectory is deterministic and can be modelled from sub-frame measurements such as those presented here, each timestamped electron could then be assigned to the probe position corresponding to its arrival time, rather than to a finite temporal bin. Because no sub-images need be formed, this avoids the signal-to-noise penalty associated with finer binning. This requires only that the scan and detector clocks share a common time origin and each detected electron can then be assigned the corresponding probe position from the fitted trajectory.

A second avenue concerns the relationship between the correction and downstream reconstruction. As implemented, the alignment is performed on per-sub-frame images and is therefore tied to a specific reconstruction mode. A natural extension is to incorporate the per-sub-frame shifts directly into ptychographic reconstructions, where the early sub-frames provide effective probe positions intermediate to those of the commanded scan grid. Together, these directions point toward a pipeline in which faster detectors and joint reconstruction methods systematically address intra-dwell artefacts at both the acquisition and reconstruction stages.

# 5 Conclusion

In summary, we have shown that scan-coil settling produces a systematic, directional smearing of 4D-STEM data at microsecond dwell times, and that this smearing can be measured directly from the sub-frame images and corrected by phase-correlation alignment in the scan-time domain. The correction is broadly applicable across scan strategies, step sizes, and probe-defocus conditions, and recovers a substantial fraction of the signal that would otherwise be lost in both atomic-resolution and low-dose imaging regimes. As fast pixelated detectors continue to push 4D-STEM toward conventional STEM speeds, intra-dwell scan-coil effects will become increasingly important to characterize and correct. The methodology presented here provides a practical, software-only route to such correction and is directly applicable to existing data and existing instruments.

## CRediT authorship contribution statement

V.K. conceived the project, performed all experiments, analysed the data and wrote the manuscript. A.J. and T.L. performed the data acquisition and analysis on LMU Munich microscope under the supervision of K.M.-C. J.R. maintained the EPFL microscope. M.K. and H.S. supervised the project. All authors contributed to the manuscript.

## Declaration of competing interest

H.S. is a consultant to Dectris. All other authors declare no competing interests.

## Acknowledgement

This work was in part supported by the Swiss National Science Foundation (grant 200021_200628), and by the European Union (ERC 4D-BioSTEM, No 101118656). Views and opinions expressed are, however, those of the authors only and do not necessarily reflect those of the European Union or the European Research Council Executive Agency. Neither the European Union nor the granting authority can be held responsible for them. We thank Gaël Cartier-Michaud from the EPFL for expert IT support.

## Appendix. Supplementary materials

Download: PDF document.

## Data availability

The raw 4D-STEM datasets generated in this work are available at the Zenodo (DOI: https://doi.org/10.5281/zenodo.21886705). The sub-frame alignment script is available as a standalone Python tool at Zenodo (DOI: https://doi.org/10.5281/zenodo.21887212).

# References

[1] D.A. Muller, Structure and bonding at the atomic scale by scanning transmission electron microscopy, Nature Materials 8 (2009) 263–270.

[2] C. Ophus, Quantitative scanning transmission electron microscopy for materials science: Imaging, diffraction, spectroscopy, and tomography, Annual Review of Materials Research 53 (2023) 105–141.

[3] S.J. Pennycook, P.D. Nellist, Scanning transmission electron microscopy: imaging and analysis, Springer Science & Business Media, 2011.

[4] V. Kumar, J.P. Camden, Imaging vibrational excitations in the electron microscope, The Journal of Physical Chemistry C 126 (2022) 16919–16927.

[5] S. Pennycook, D. Jesson, High-resolution Z-contrast imaging of crystals, Ultramicroscopy 37 (1991) 14–38.

[6] P.D. Nellist, S.J. Pennycook, The principles and interpretation of annular dark-field Z-contrast imaging, in: Advances in Imaging and Electron Physics, Elsevier, 2000: pp. 147–203.

[7] S. Toyama, T. Seki, Y. Kohno, Y.O. Murakami, Y. Ikuhara, N. Shibata, Nanoscale electromagnetic field imaging by advanced differential phase-contrast STEM, Nature Reviews Electrical Engineering 2 (2025) 27–41.

[8] N. Shibata, S.D. Findlay, Y. Kohno, H. Sawada, Y. Kondo, Y. Ikuhara, Differential phase-contrast microscopy at atomic resolution, Nature Physics 8 (2012) 611–615.

[9] C. Ophus, Four-dimensional scanning transmission electron microscopy (4D-STEM): From scanning nanodiffraction to ptychography and beyond, Microscopy and Microanalysis 25 (2019) 563–582.

[10] Z. Chen, Y. Jiang, Y.-T. Shao, M.E. Holtz, M. Odstrčil, M. Guizar-Sicairos, I. Hanke, S. Ganschow, D.G. Schlom, D.A. Muller, Electron ptychography achieves atomic-resolution limits set by lattice vibrations, Science 372 (2021) 826–831.

[11] D. Ma, G. Li, D.A. Muller, S.E. Zeltmann, Information in 4D-STEM: Where it is, and how to use it, Ultramicroscopy (2026) 114351.

[12] K.C. Bustillo, S.E. Zeltmann, M. Chen, J. Donohue, J. Ciston, C. Ophus, A.M. Minor, 4D-STEM of beam-sensitive materials, Accounts of Chemical Research 54 (2021) 2543–2551.

[13] G. Li, H. Zhang, Y. Han, 4D-STEM ptychography for electron-beam-sensitive materials, ACS Central Science 8 (2022) 1579–1588.

[14] G. Li, M. Xu, W.-Q. Tang, Y. Liu, C. Chen, D. Zhang, L. Liu, S. Ning, H. Zhang, Z.-Y. Gu, Atomically resolved imaging of radiation-sensitive metal-organic frameworks via electron ptychography, Nature Communications 16 (2025) 914–914.

[15] B. Küçükoğlu, I. Mohammed, R.C. Guerrero-Ferreira, S.M. Ribet, G. Varnavides, M.L. Leidl, K. Lau, S. Nazarov, A. Myasnikov, M. Kube, J. Radecke, C. Sachse, K. Müller-Caspary, C. Ophus, H. Stahlberg, Low-dose cryo-electron ptychography of proteins at sub-nanometer resolution, Nature Communications 15 (2024) 8062–8062. https://doi.org/10.1038/s41467-024-52403-5.

[16] Y. Yu, K.A. Spoth, M. Colletta, K.X. Nguyen, S.E. Zeltmann, X.S. Zhang, M. Paraan, M. Kopylov, C. Dubbeldam, D. Serwas, Dose-efficient cryo-electron microscopy for thick samples using tilt-corrected scanning transmission electron microscopy, Nature Methods 22 (2025) 2138–2148.

[17] J. Zivanov, T. Nakane, S.H.W. Scheres, A Bayesian approach to beam-induced motion correction in cryo-EM single-particle analysis, IUCrJ 6 (2019) 5–17. https://doi.org/10.1107/S205225251801463X.

[18] A. Punjani, J.L. Rubinstein, D.J. Fleet, M.A. Brubaker, cryoSPARC: algorithms for rapid unsupervised cryo-EM structure determination, Nature Methods 14 (2017) 290–296. https://doi.org/10.1038/nmeth.4169.

[19] T. Grant, N. Grigorieff, Measuring the optimal exposure for single particle cryo-EM using a 2.6 Å reconstruction of rotavirus VP6, Elife 4 (2015) e06980–e06980.

[20] D.G. Stroppa, M. Meffert, C. Hoermann, P. Zambon, D. Bachevskaya, H. Remigy, C. Schulze-Briese, L. Piazza, From STEM to 4D STEM: Ultrafast diffraction mapping with a hybrid-pixel detector, Microscopy Today 31 (2023) 10–14.

[21] D. Jannis, C. Hofer, C. Gao, X. Xie, A. Béché, T.J. Pennycook, J. Verbeeck, Event driven 4D STEM acquisition with a Timepix3 detector: Microsecond dwell time and faster scans for high precision and low dose applications, Ultramicroscopy 233 (2022) 113423–113423.

[22] T. Poikela, J. Plosila, T. Westerlund, M. Campbell, M.D. Gaspari, X. Llopart, V. Gromov, R. Kluit, M.V. Beuzekom, F. Zappon, Timepix3: a 65K channel hybrid pixel readout chip with simultaneous ToA/ToT and sparse readout, Journal of Instrumentation 9 (2014) C05013–C05013.

[23] J.P. Buban, Q. Ramasse, B. Gipson, N.D. Browning, H. Stahlberg, High-resolution low-dose scanning transmission electron microscopy, Journal of Electron Microscopy 59 (2010) 103–112.

[24] S. Ning, T. Fujita, A. Nie, Z. Wang, X. Xu, J. Chen, M. Chen, S. Yao, T.-Y. Zhang, Scanning distortion correction in STEM images, Ultramicroscopy 184 (2018) 274–283.

[25] T. Mullarkey, J.J.P. Peters, C. Downing, L. Jones, Using your beam efficiently: Reducing electron dose in the STEM via flyback compensation, Microscopy and Microanalysis 28 (2022) 1428–1436.

[26] C. Ophus, J. Ciston, C.T. Nelson, Correcting nonlinear drift distortion of scanning probe and scanning transmission electron microscopies from image pairs with orthogonal scan directions, Ultramicroscopy 162 (2016) 1–9.

[27] L. Jones, P.D. Nellist, Identifying and correcting scan noise and drift in the scanning transmission electron microscope, Microscopy and Microanalysis 19 (2013) 1050–1060.

[28] X. Sang, A.R. Lupini, R.R. Unocic, M. Chi, A.Y. Borisevich, S.V. Kalinin, E. Endeve, R.K. Archibald, S. Jesse, Dynamic scan control in STEM: spiral scans, Advanced Structural and Chemical Imaging 2 (2016) 6–6.

[29] M. Palos, L. Spillane, G. Topore, Y. Li, D. Pesquera, C. Ophus, S.M. Ribet, M.S. Conroy, Programmable Beam Control for Electron Energy-Loss Spectroscopy and Ptychography, arXiv Preprint arXiv:2509.10726 (2025).

[30] S. Ning, W. Xu, L. Loh, Z. Lu, M. Bosman, F. Zhang, Q. He, An integrated constrained gradient descent (iCGD) protocol to correct scan-positional errors for electron ptychography with high accuracy and precision, Ultramicroscopy 248 (2023) 113716–113716.

[31] M. Odstrčil, A. Menzel, M. Guizar-Sicairos, Iterative least-squares solver for generalized maximum-likelihood ptychography, Optics Express 26 (2018) 3108–3123.

[32] V. Kumar, H. Stahlberg, M. Kube, Dynamic Response of the Beam Deflection System for High-Speed 4D-STEM, Microscopy and Microanalysis 32 (2026) ozag053-049.

[33] R. Egoavil, P. Potocek, E. Van Cappellen, M. Meledina, M. Wirix, B. Freitag, Advantages of an Electrostatic Beam Blanker in STEM, Microscopy and Microanalysis 31 (2025) ozaf048-104.

[34] V. Kumar, J. Radecke, C. K.V., I. Mohammed, R.C. Guerrero-Ferreira, D. Harder, D. Fotiadis, H. Stahlberg, Pulsed-electron illumination does not reduce beam damage for imaging biological macromolecules, Nature Communications (2026). https://doi.org/10.1038/s41467-026-74316-1.

[35] G. Bongiovanni, M.M. van Rijt, O. Shánĕl, E.R. Kieft, Advancing ultrafast (S) TEM with the combination of an RF cavity and an electrostatic beam blanker, Structural Dynamics 12 (2025).

[36] B.H. Savitzky, S.E. Zeltmann, L.A. Hughes, H.G. Brown, S. Zhao, P.M. Pelz, T.C. Pekin, E.S. Barnard, J. Donohue, L. Rangel DaCosta, py4DSTEM: A software package for four-dimensional scanning transmission electron microscopy data analysis, Microscopy and Microanalysis 27 (2021) 712–743.

[37] S. Van der Walt, J.L. Schönberger, J. Nunez-Iglesias, F. Boulogne, J.D. Warner, N. Yager, E. Gouillart, T. Yu, scikit-image: image processing in Python, PeerJ 2 (2014) e453–e453.

[38] G. Varnavides, S.M. Ribet, S.E. Zeltmann, Y. Yu, B.H. Savitzky, D.O. Byrne, F.I. Allen, V.P. Dravid, M.C. Scott, C. Ophus, Iterative phase retrieval algorithms for scanning transmission electron microscopy, arXiv Preprint arXiv:2309.05250 (2023).

[39] G. Varnavides, J.M. Bekkevold, S.M. Ribet, M.C. Scott, L. Jones, C. Ophus, Relaxing direct ptychography sampling requirements via parallax imaging insights, Microscopy and Microanalysis 32 (2026) ozaf139.

[40] Y. Cheng, N. Grigorieff, P.A. Penczek, T. Walz, A primer to single-particle cryo-electron microscopy, Cell 161 (2015) 438–449.

[41] A. Béché, B. Goris, B. Freitag, J. Verbeeck, Development of a fast electromagnetic beam blanker for compressed sensing in scanning transmission electron microscopy, Applied Physics Letters 108 (2016).

# Supporting Information

## Scan-Coil Delay Causes Anisotropic Signal Loss in Fast 4D-STEM

Vishal Kumar[1,2,#], Andreas Jehle,[3] Tizian Lorenzen,[3] Julika Radecke[1,2], Knut Müller-Caspary,[3] Massimo Kube[1,2], and Henning Stahlberg[1,2,#]

1. Laboratory of Biological Electron Microscopy, Institute of Physics, School of Basic Sciences, EPFL, Rt. de la Sorge, 1015 Lausanne, Switzerland
2. Department of Fundamental Microbiology, Faculty of Biology and Medicine, University of Lausanne, Rt. de la Sorge, 1015 Lausanne, Switzerland.
3. Department of Chemistry and Center for NanoScience, Ludwig-Maximilians-Universität München, Butenandtstr. 11, 81377 München, Germany

\# Corresponding authors: Vishal.Kumar@epfl.ch, Henning.Stahlberg@epfl.ch

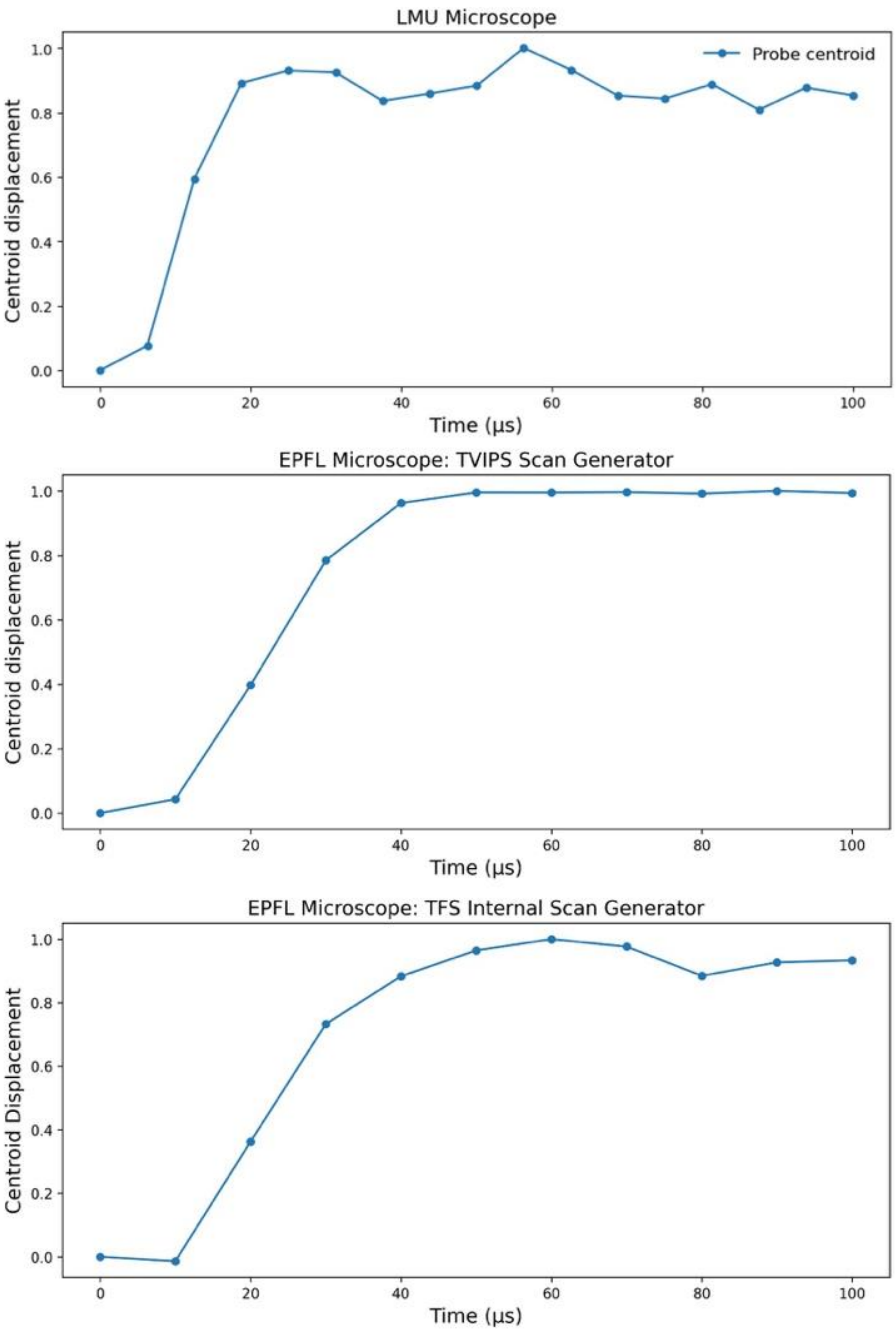

**Supplementary Figure 1.** Intra-dwell probe settling is common across instruments and scan generators. Normalized probe centroid displacement versus intra-dwell time for the (top) LMU microscope, (middle) EPFL microscope with the TVIPS Universal Scan Generator, and (bottom) the TFS internal scan generator. Each curve is normalized to its settled (plateau) value; in all three cases the centroid rises from the previous commanded position and settles

within a few tens of microseconds, confirming that the scan-coil settling is a generic feature of the scan hardware rather than of any single instrument or scan generator.

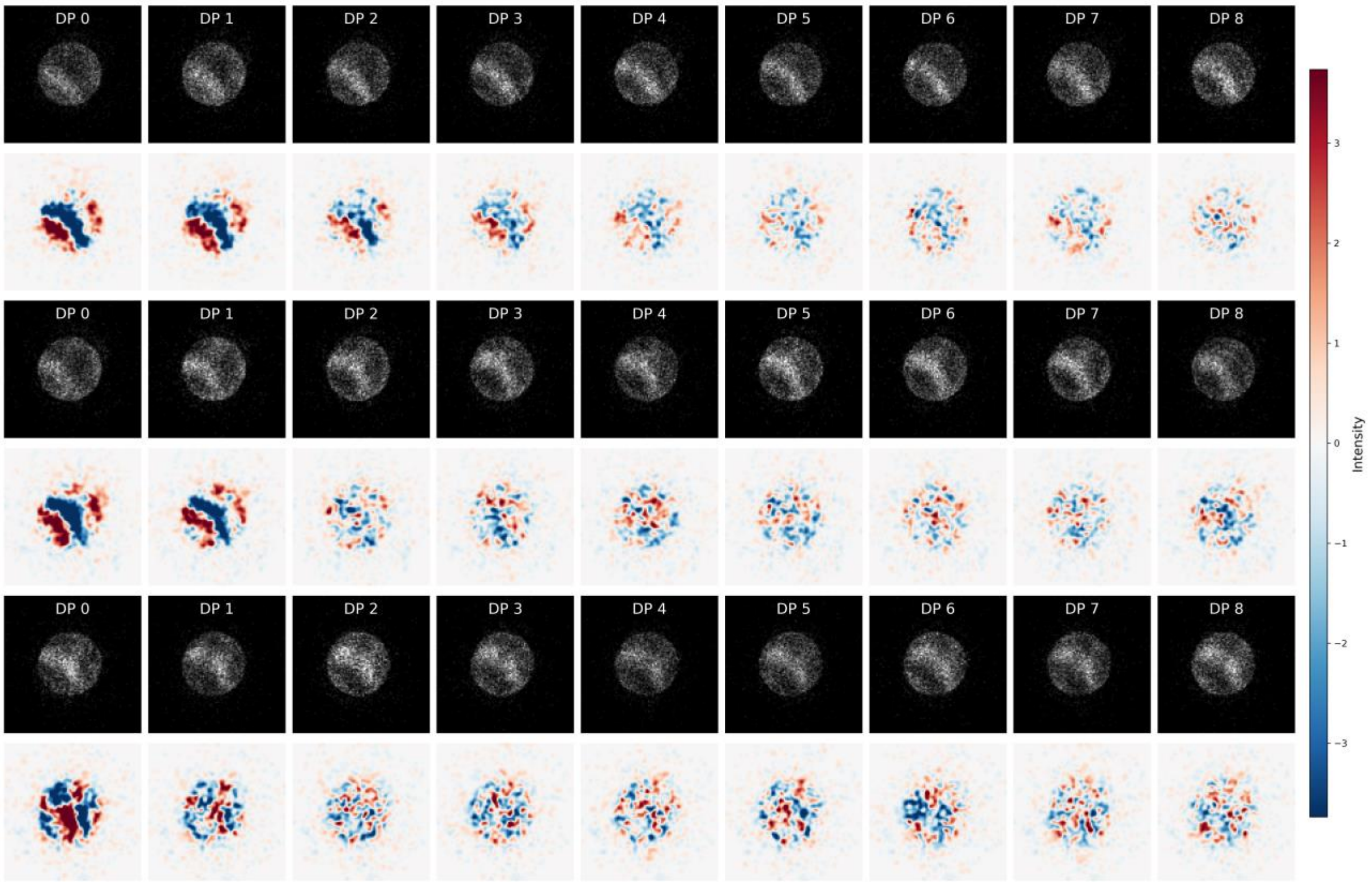

**Supplementary Figure 2.** Intra-dwell diffraction-space settling is consistent across the field of view. Each pair of rows shows a different probe (scan) position: nine consecutive 10 μs sub-frame diffraction patterns (DP 0–DP 8, top of each pair) and their differences from the final sub-frame (bottom, Gaussian-smoothed with σ = 1.2 pixels for clarity). At every position the bright-field disc shows a systematic intensity redistribution in the early sub-frames that decays into noise by roughly the fourth sub-frame, mirroring the real-space probe settling.

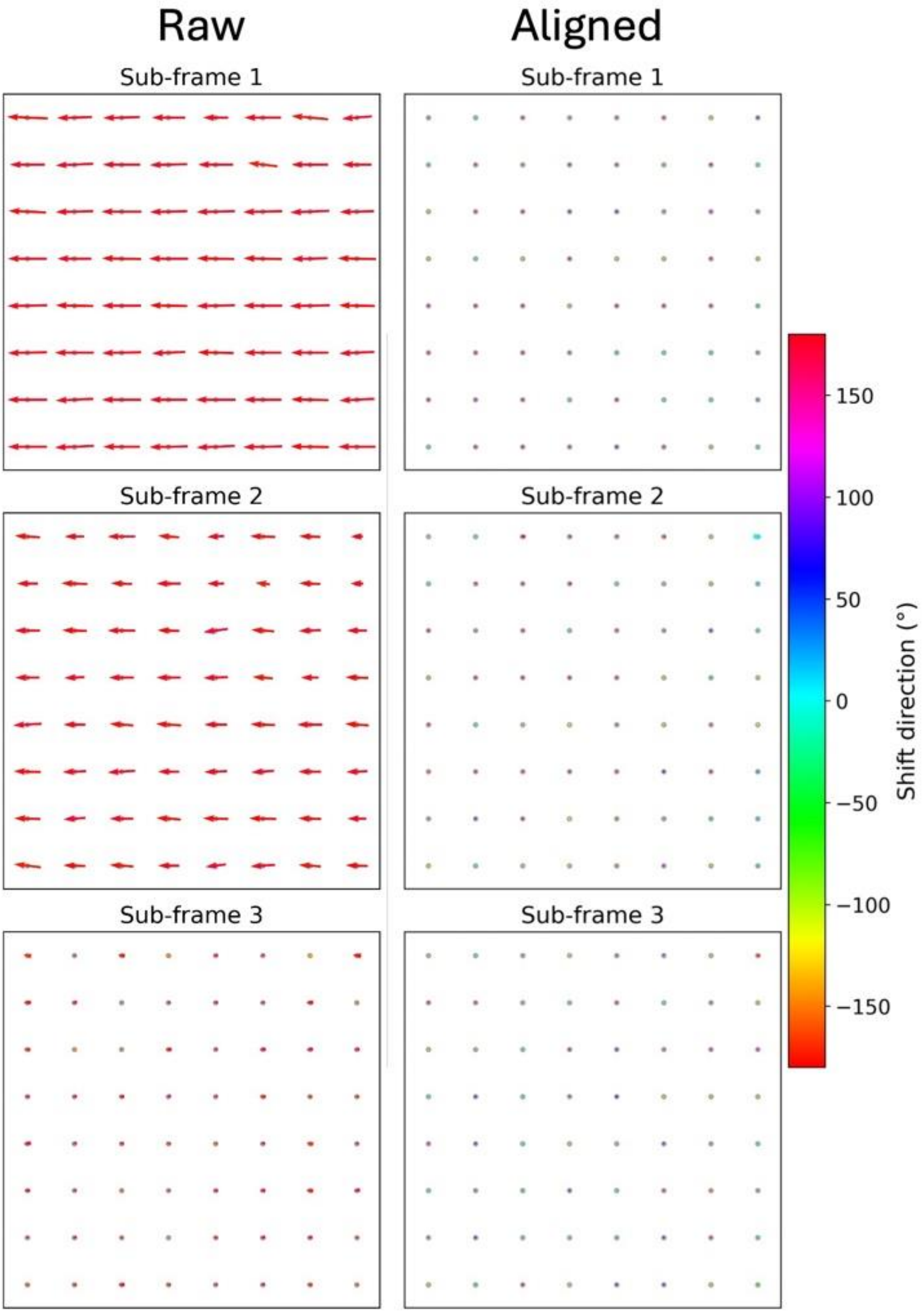

**Supplementary Figure 3.** Per-position misalignment before and after correction (for raster scan). Local shift vectors of the per-sub-frame vDF images relative to the final (settled) sub-frame, measured across the field of view for sub-frames 1, 2, and 3, shown before (Raw, left) and after (Aligned, right) the phase-correlation correction. Arrow length encodes shift magnitude and colour encodes shift direction. In the raw data the shifts are large and uniformly

oriented along the fast-scan direction. After correction the residual shifts collapse to near zero at all positions, confirming that a single per-sub-frame shift registers the whole field of view.

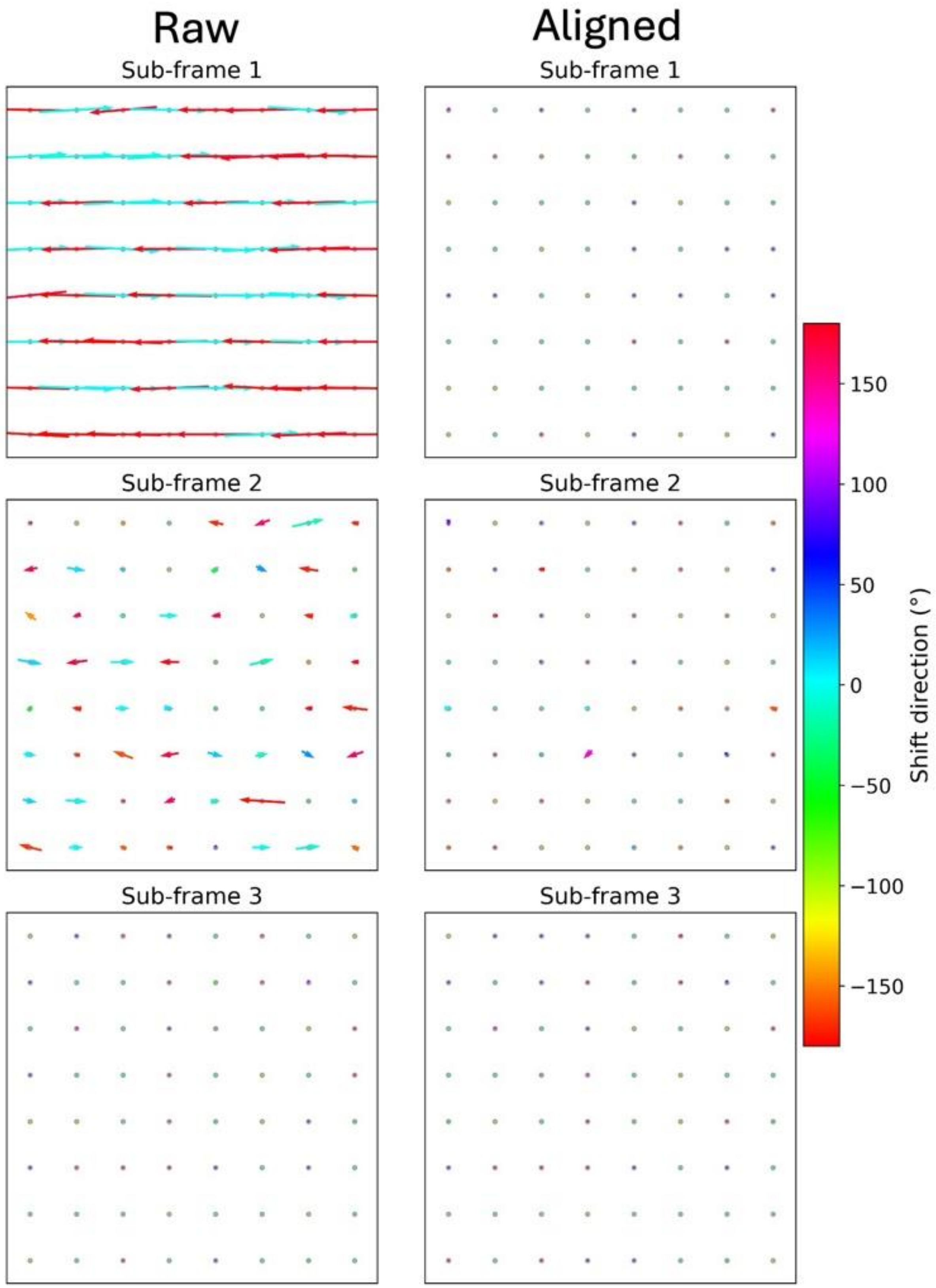

**Supplementary Figure 4.** Measured shifts follow the scan trajectory (serpentine scan). Local shift vectors of the per-sub-frame vDF images relative to the final sub-frame, across the field of view for sub-frames 1, 2, and 3, before (Raw, left) and after (Aligned, right) correction. Arrow length encodes shift magnitude and colour encodes shift direction. In the raw data the shifts are large and uniformly oriented along the fast-scan direction. After correction the residual shifts collapse to near zero at all positions, confirming that a single per-sub-frame shift registers the whole field of view.

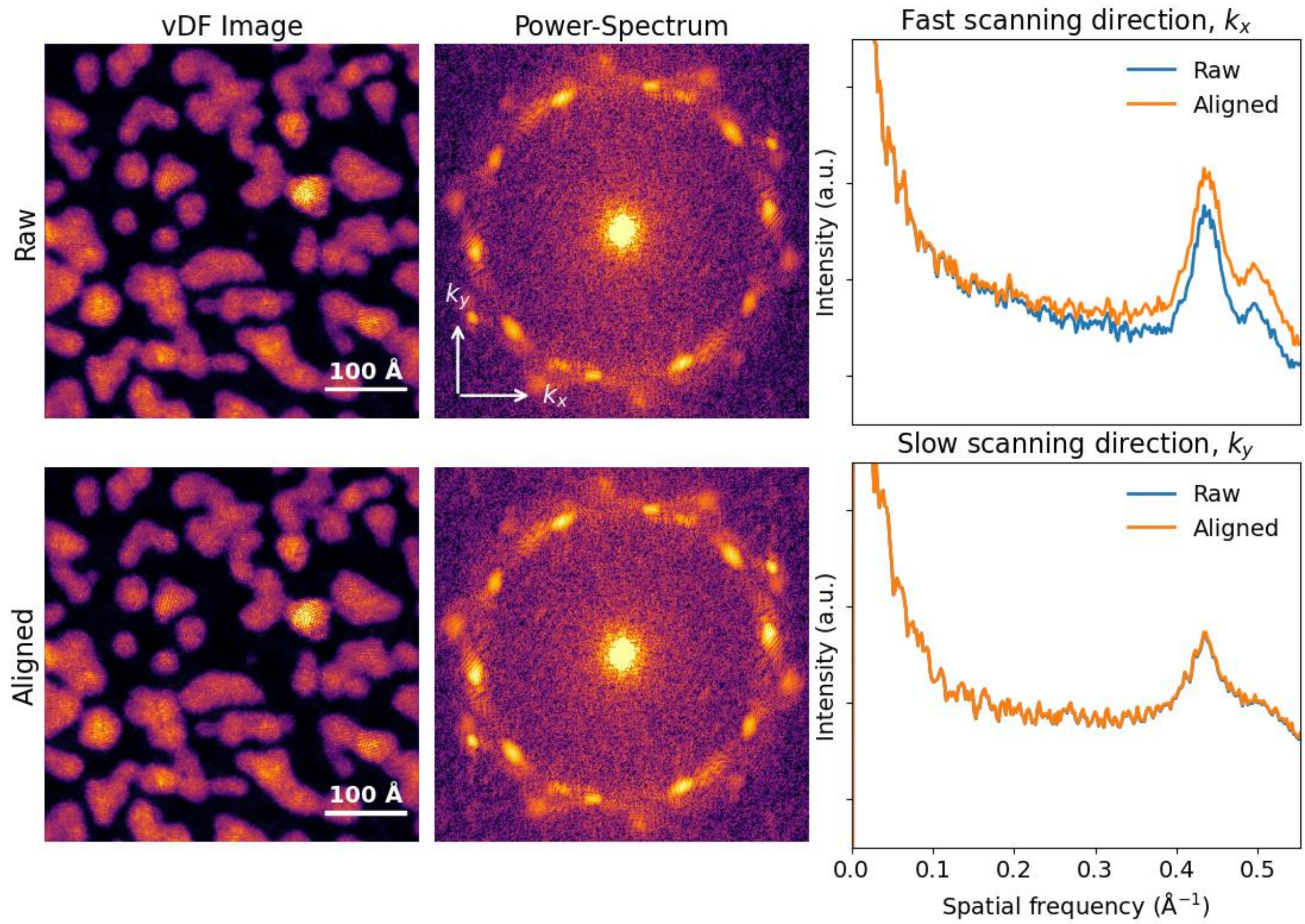

**Supplementary Figure 5.** Raw and sub-frame-aligned vDF images at a step size of 0.905 Å, with the corresponding power spectra and directional radial profiles. Raw and aligned panels share identical display limits, and profiles are averaged over ±20° wedges about each axis. The images and power spectra are visually indistinguishable because the residual displacement at this step size is only about one scan pixel, far below the threshold of visual detection in a full-field display. The radial profiles resolve it unambiguously. Along the fast-scan direction, sub-frame alignment recovers significant power at the gold lattice spacings, $k = 0.43$ Å$^{-1}$ (Au(111), $d = 2.35$ Å) and $k = 0.50$ Å$^{-1}$ (Au(200), $d = 2.04$ Å). Along the slow-scan direction, where no coil settling transient acts, the raw and aligned profiles are indistinguishable across the full frequency range.

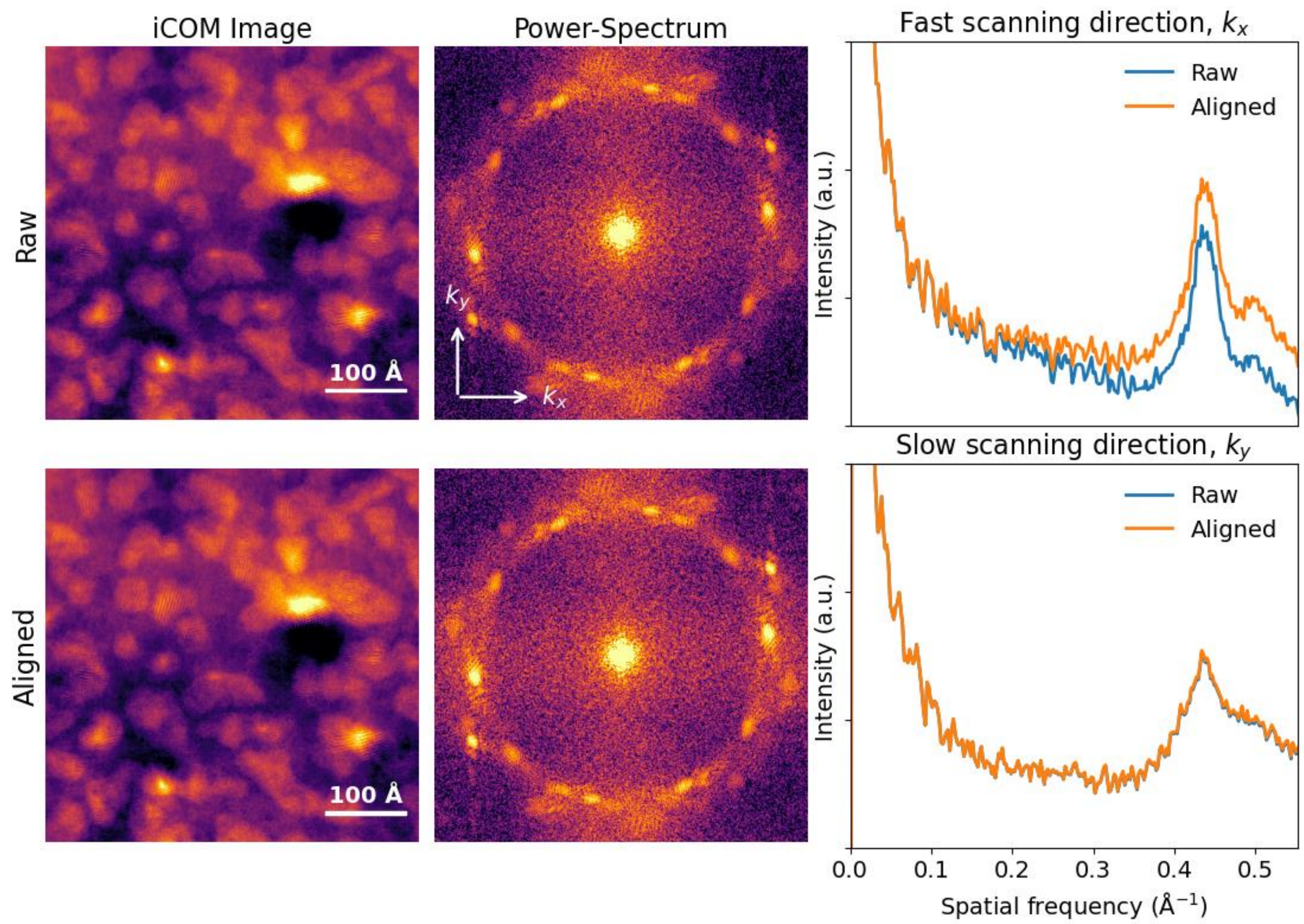

**Supplementary Figure 6.** Raw and sub-frame-aligned iCoM images at a step size of 0.905 Å, with the corresponding power spectra and directional radial profiles. Raw and aligned panels share identical display limits, and profiles are averaged over ±20° wedges about each axis. The images and power spectra are visually indistinguishable because the residual displacement at this step size is only about one scan pixel, far below the threshold of visual detection in a full-field display. The radial profiles resolve it unambiguously. Along the fast-scan direction, sub-frame alignment recovers significant power at the gold lattice spacings, k = 0.43 Å$^{-1}$ (Au(111), d = 2.35 Å) and k = 0.50 Å$^{-1}$ (Au(200), d = 2.04 Å). Along the slow-scan direction, where no coil settling transient acts, the raw and aligned profiles are indistinguishable across the full frequency range.

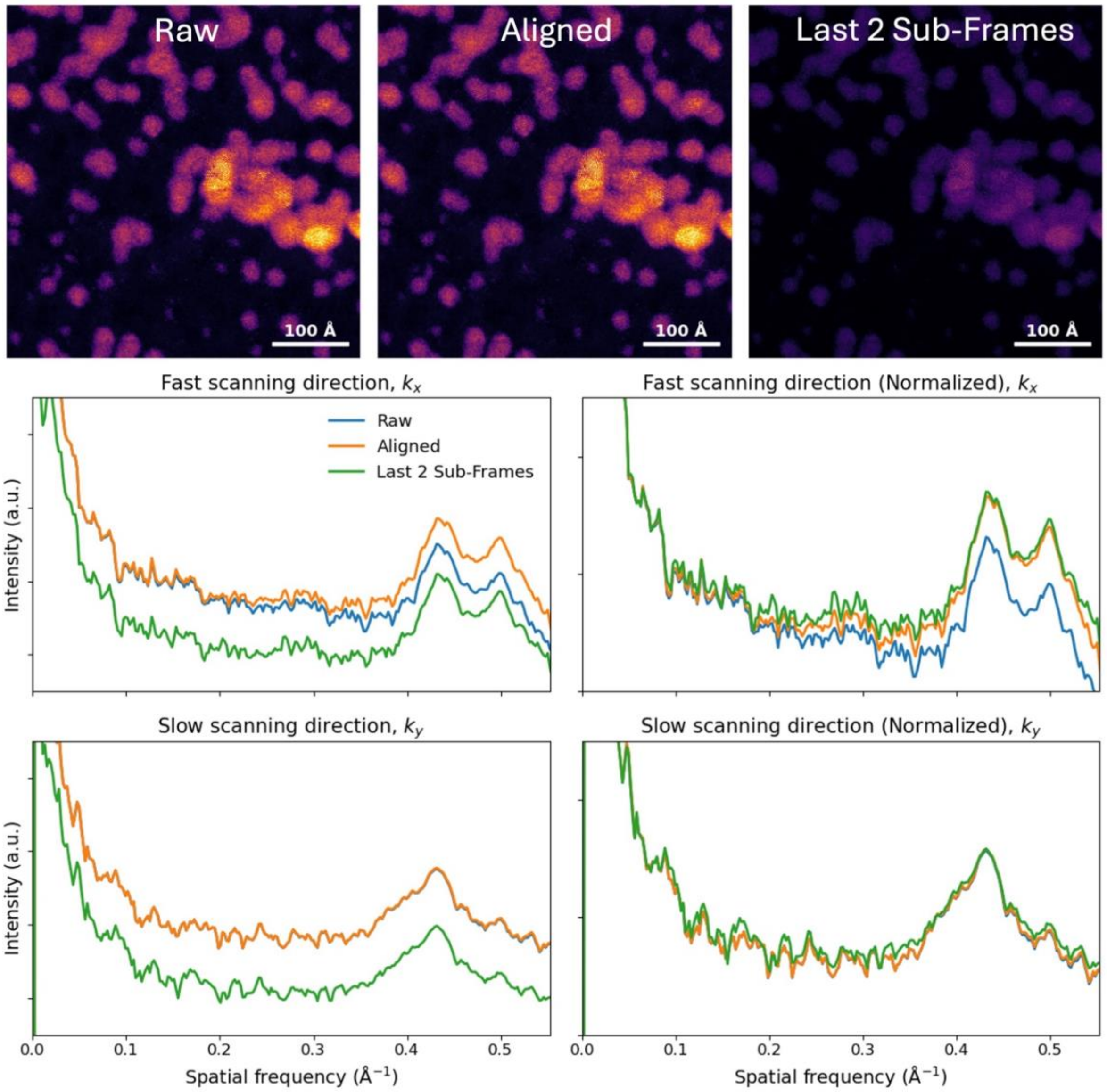

**Supplementary Figure 7.** Comparison of different reconstruction strategies at a 0.905 Å scan step. (a) vDF images formed from the raw sub-frames (1–5), aligned sub-frames (1-5), and from the two settled sub-frames (4–5). All three panels share identical intensity limits, so the reduced brightness of the third reflects its lower dose (40%) rather than any loss of structure. Radial power profiles taken in about the fast-scan ($k_x$) and slow-scan ($k_y$) directions for (b) as acquired and (c) normalized images, which removes the scaling with total dose and allows the contrast per electron to be compared directly. For normalized profiles along the fast-scan direction, sub-frame alignment and settled sub-frames (subframe 4-5) recover the same power at the Au(111) and Au(200) spacings, confirming that the power loss in the raw image originates from intra-dwell smearing rather than from any other source. Along the slow-scan direction the three are indistinguishable for normalized images. The lower profiles of the settled sub-frames in the as-acquired panels reflect only their reduced dose, leading to lower SNR (0.63 or √0.4) compared to the aligned image.

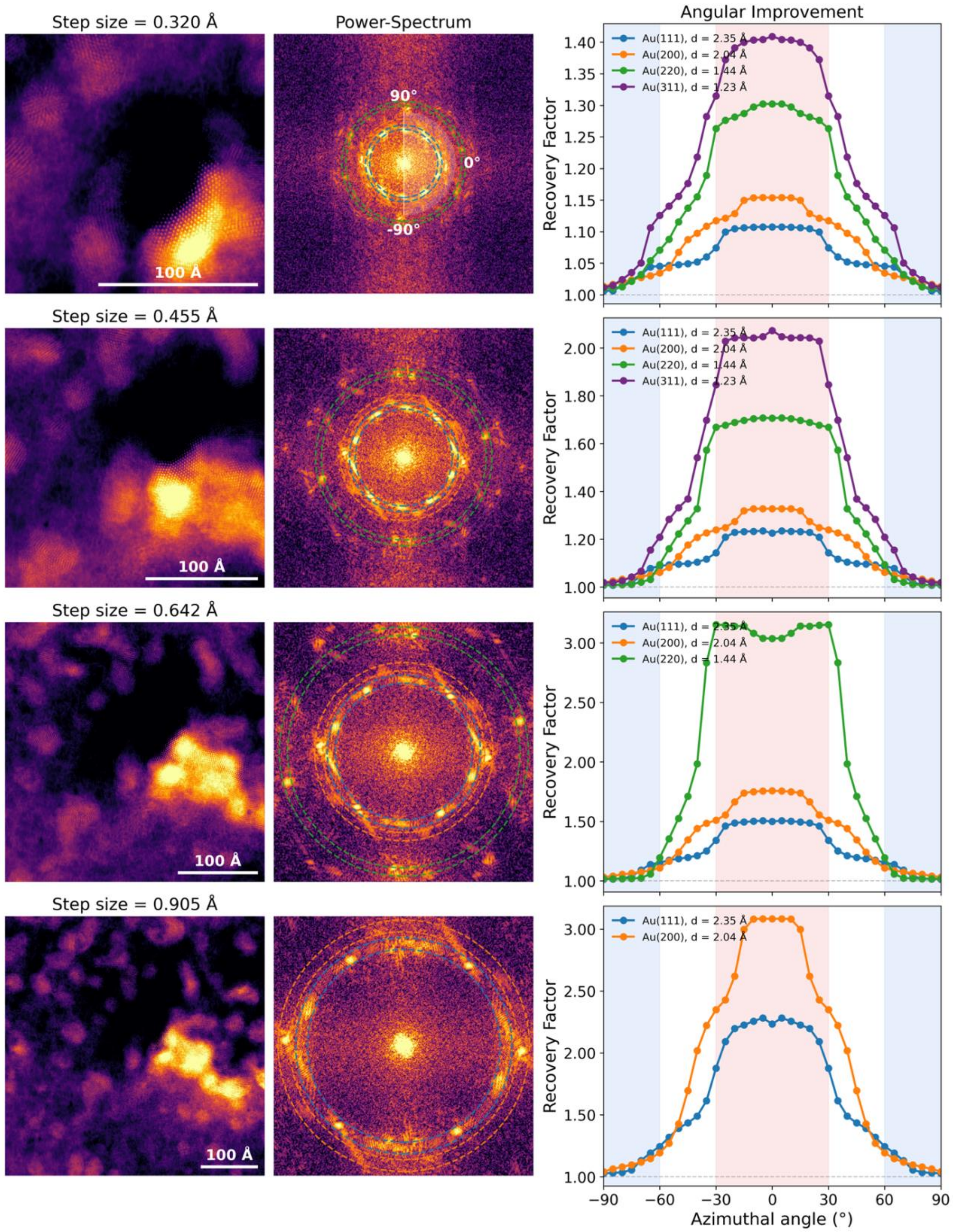

**Supplementary Figure 8.** Step-size and frequency dependence of the alignment correction for iCoM reconstructions. iCoM images, power spectra, and recovery factor versus azimuthal angle for four scan step sizes: 0.32 Å, 0.455 Å, 0.642 Å, and 0.905 Å. Dashed circles mark integration windows used to compute the recovery factor at each accessible Au reflection (d = 2.35, 2.04, 1.44, 1.23 Å for Au(111), (200), (220), (311)). Pink and blue shaded regions in the azimuthal plots mark the fast-scanning (±30°) and slow-scanning (±60–90°) directions. Higher-order reflections become inaccessible at larger step sizes due to Nyquist sampling.

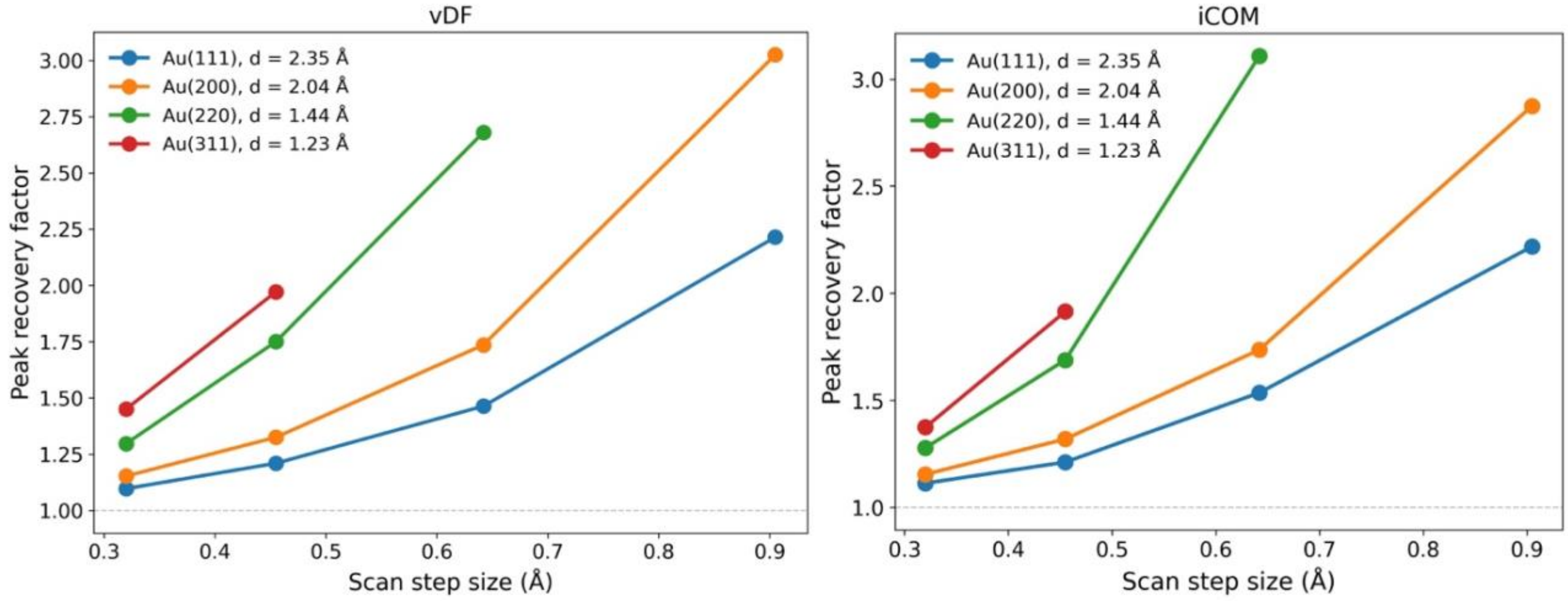

**Supplementary Figure 9.** Peak recovery factor versus scan step size for (left) vDF and (right) iCoM reconstructions. Each curve shows the peak (scanning-direction) recovery factor at a given Au reflection (Au(111), (200), (220), (311); d = 2.35, 2.04, 1.44, 1.23 Å) as a function of scan step size (0.32–0.905 Å). For both reconstruction modes the recovery factor increases monotonically with step size and is systematically larger for higher-order reflections, confirming that the intra-dwell smearing acts as an effective probe broadening along the fast-scan direction that suppresses finer spatial features more strongly. Higher-order reflections are accessible only at the smaller step sizes due to Nyquist sampling.

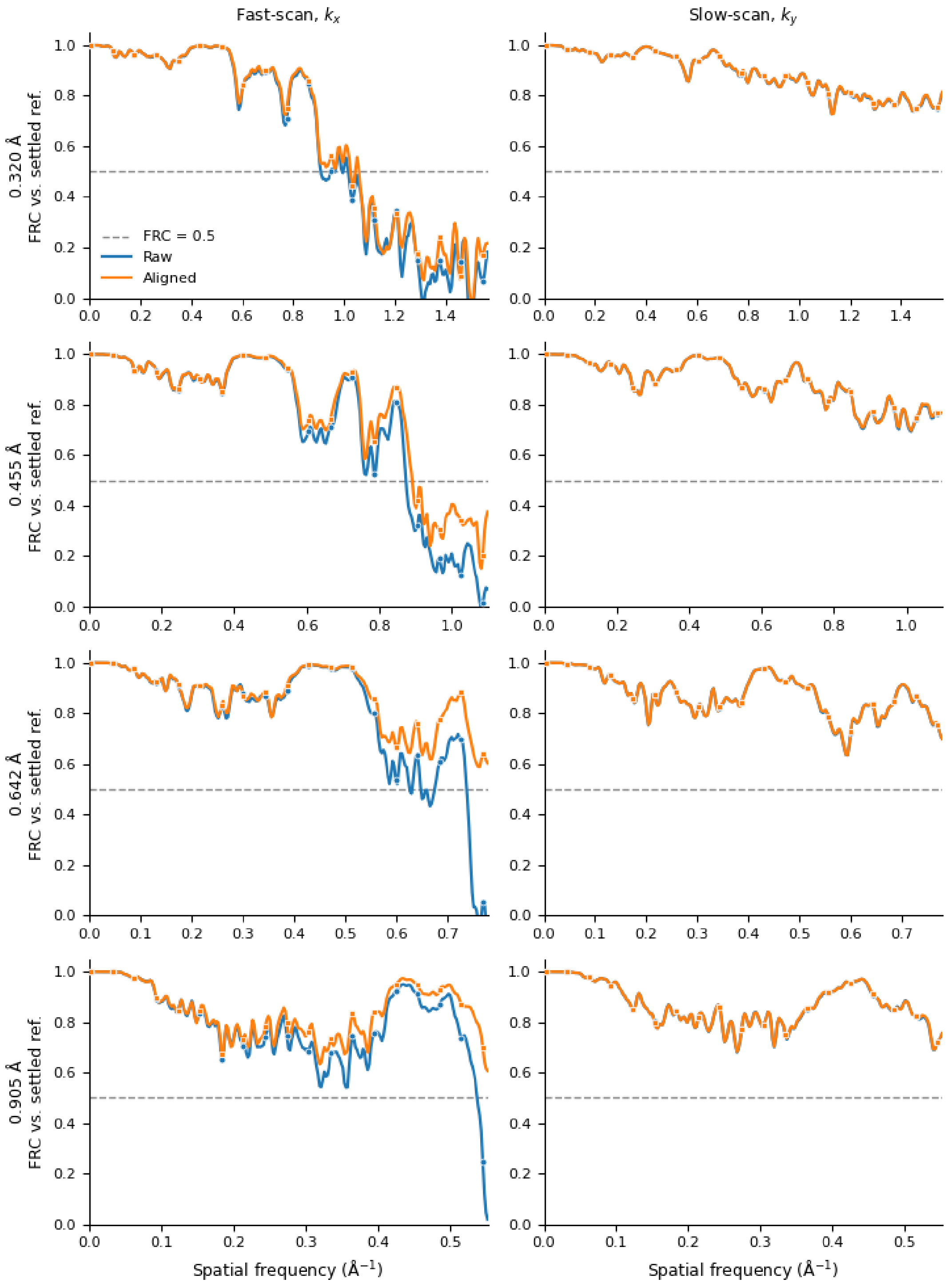

**Supplementary Figure 10.** Directional Fourier ring correlation (FRC) for vDF reconstructions at different scan step sizes. Each row is one step size; left, a ±20° wedge about the fast-scan direction ($k_x$); right, the same about the slow-scan direction ($k_y$). The test images are the sum of sub-frames 1–4 before (blue) and after (orange) sub-frame alignment; the reference is sub-frame 5 (40–50 µs), by which time the probe has settled. Alignment raises the fast-scan correlation, while the slow-scan curves are unchanged throughout, confirming that the correction acts along the fast-scan direction only.

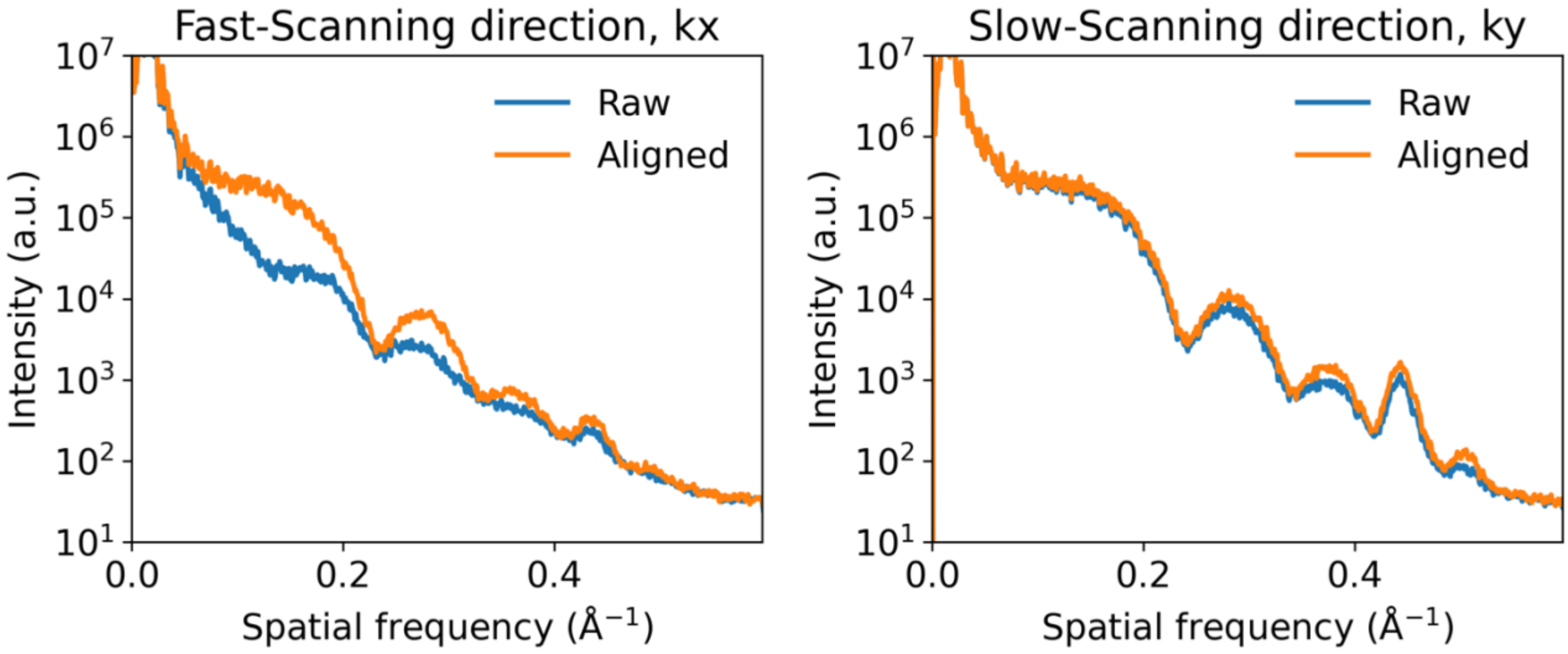

**Supplementary Figure 11.** Directional recovery in defocused parallax imaging at a 5.025 Å scan step. Radial power-spectrum profiles (intensity versus spatial frequency) of the raw and aligned parallax reconstructions in the (left) fast-scanning direction ($k_x$) and (right) slow-scanning direction ($k_y$), each averaged over a ±20° wedge. In the fast-scanning direction the aligned profile lies systematically above the raw one over a broad frequency band, showing recovered signal, whereas in the slow-scanning direction the two profiles overlap, confirming the smearing and its correction act only along the fast-scan direction.

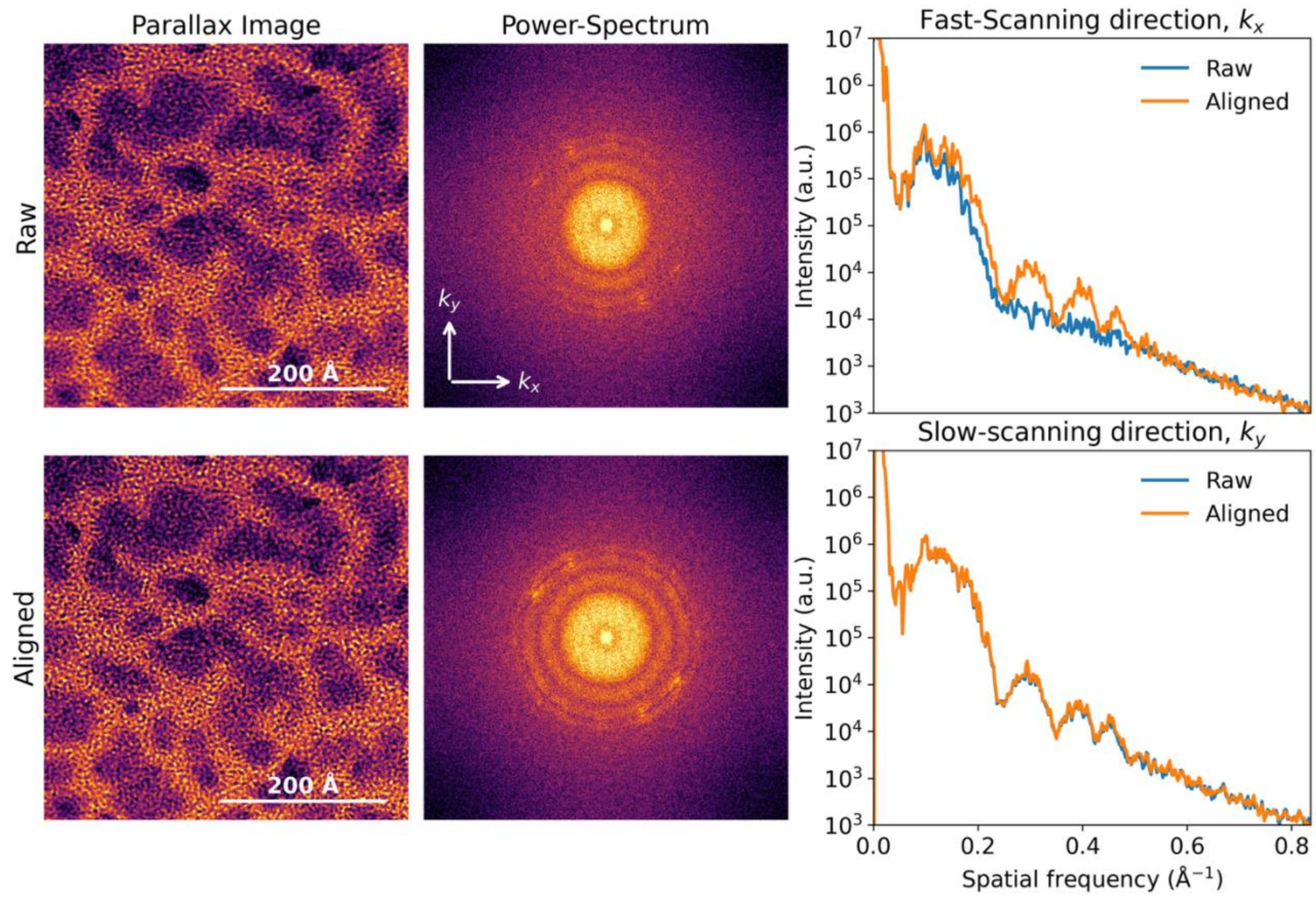

**Supplementary Figure 12.** Broadband parallax recovery on the data from Titan Themis at LMU microscope at a 2.16 Å scan step (defocus ≈ 60 nm). (left) Raw (top) and aligned (bottom) parallax images; (middle) their corresponding power spectra, with $k_x$ (fast-scanning) and $k_y$ (slow-scanning) directions indicated; (right) radial power-spectrum profiles of the raw and aligned reconstructions in the fast-scanning ($k_x$, top) and slow-scanning ($k_y$, bottom) directions. After alignment, more information is recovered along the fast-scanning direction: the aligned profile rises above the raw one over a broad frequency band and the lattice spots become more intense.

**Supplementary Table 1.** Microscope and Scan-generator Specifications.

| Microscope Name | Location | Scan-Generator |
|---|---|---|
| Titan Krios G4 | EPFL, Lausanne, Switzerland | TVIPS USG, and ThermoFisher Internal Scan-Generator |
| Titan Themis | LMU, Munich, Germany | DigiScan from Gatan |

**Supplementary Table 2.** Upsampling factors used for the parallax reconstructions at each scan step size.

| Microscope Location | Step-size (Å) | Upsampling Factor |
|---|---|---|
| EPFL, Lausanne, Switzerland | 0.91 | 1.5 |
| | 1.285 | 2 |
| | 2.525 | 3 |
| | 5.025 | 6 |
| LMU, Munich, Germany | 2.154 | 4 |